\documentclass[12pt,twoside,a4paper]{article}
\usepackage{amsmath}
\usepackage{amssymb}
\usepackage{axodraw}
\usepackage{epsfig}
\def\msbar{\ensuremath{{\rm{\overline{MS}}}}} 
\newcommand{\msbarm}{\ensuremath{\msbar_m}} 
\newcommand{\pbar}{\ensuremath{\bar{p}}}
\newcommand{\qbar}{\ensuremath{\bar{q}}}
\newcommand{\Qbar}{\ensuremath{\overline{Q}}}
\newcommand{\Ord}{\ensuremath{{\cal O}}}
\newcommand{\der}{\ensuremath{{\operatorname{d}}}}
\newcommand{\sub}{\ensuremath{\mathrm{sub}}}
\newcommand{\sigmahat}{\ensuremath{\hat{\sigma}}}
\newcommand{\mui}{\ensuremath{\mu_F}}
\newcommand{\muis}{\ensuremath{\mu_F^2}}
\newcommand{\muf}{\ensuremath{\mu_F^\prime}}
\newcommand{\mufs}{\ensuremath{\mu_F^{\prime \,2}}}
\newcommand{\mur}{\ensuremath{\mu_R}}
\newcommand{\murs}{\ensuremath{\mu_R^2}}
\newcommand{\shat}{\ensuremath{\hat{s}}}
\newcommand{\vhat}{\ensuremath{\hat{v}}}
\newcommand{\what}{\ensuremath{\hat{w}}}
\newcommand{\that}{\ensuremath{\hat{t}}}
\newcommand{\uhat}{\ensuremath{\hat{u}}}
\newcommand{\xb}{\ensuremath{\bar{x}}}
\newcommand{\zb}{\ensuremath{\bar{z}}}
\newcommand{\dcc}{\ensuremath{d_{Q\to Q}^{(1)}}}
\newcommand{\dcg}{\ensuremath{d_{g \to Q}^{(1)}}}
\newcommand{\fcg}{\ensuremath{f_{g\to Q}^{(1)}}}
\newcommand{\fcc}{\ensuremath{f_{Q\to Q}^{(1)}}}
\newcommand{\fgg}{\ensuremath{f_{g\to g}^{(1)}}}
\newcommand{\CF}{\ensuremath{C_F}}
\newcommand{\Lfin}{\ensuremath{\ln \frac{{\muf}^2}{m^2}}}
\newcommand{\Lini}{\ensuremath{\ln \frac{{\mui}^2}{m^2}}}
\newcommand{\cqed}[1]{\ensuremath{c^{\rm qed}_{#1}}}
\newcommand{\coqu}[1]{\ensuremath{c^{\rm oq}_{#1}}}
\newcommand{\ckqu}[1]{\ensuremath{c^{\rm kq}_{#1}}}
\newcommand{\dcqed}[1]{\ensuremath{\Delta \cqed{#1}}}
\newcommand{\dcoqu}[1]{\ensuremath{\Delta \coqu{#1}}}
\newcommand{\dckqu}[1]{\ensuremath{\Delta \ckqu{#1}}}

\newcommand{\Cgq}{\ensuremath{C_{gq}(s)}}
\newcommand{\tauq}{\ensuremath{\tau_q(v)}}
\newcommand{\ccf}[1]{\ensuremath{c^{\rm cf}_{#1}}}
\newcommand{\cca}[1]{\ensuremath{c^{\rm ca}_{#1}}}
\newcommand{\dccf}[1]{\ensuremath{\Delta \ccf{#1}}}
\newcommand{\dcca}[1]{\ensuremath{\Delta \cca{#1}}}

\newcommand{\ggQ}{\ensuremath{g+g\to Q + \Qbar + g}}
\newcommand{\gqQ}{\ensuremath{g+q \to Q + \Qbar + q}}
\newcommand{\qqbQ}{\ensuremath{q+\qbar \to Q + \Qbar + g}}
\newcommand{\gqbQ}{\ensuremath{g+\qbar \to Q + \Qbar + \qbar}}
\newcommand{\gamgQ}{\ensuremath{\gamma+g\to Q + \Qbar + g}}
\newcommand{\gamqQ}{\ensuremath{\gamma+q \to Q + \Qbar + q}}

\begin{document} 
\thispagestyle{empty} 

\title{
\vskip-3cm
{\baselineskip14pt
\centerline{\normalsize DESY 05--030 \hfill ISSN 0418--9833}
\centerline{\normalsize MZ--TH/05--04 \hfill} 
\centerline{\normalsize hep--ph/0502194 \hfill}} 
\vskip1.8cm
{\bf Collinear Subtractions in Hadroproduction}\\
{\bf of Heavy Quarks}
\author{
{\bf B.~A.~Kniehl, G.~Kramer, I.~Schienbein}
\vspace{2mm} \\
{\em \normalsize II. Institut f\"ur Theoretische Physik, Universit\"at
  Hamburg,}
\\  
{\em \normalsize Luruper Chaussee 149, D-22761 Hamburg, Germany} 
\\
{\em \normalsize E-mail:} 
{\normalsize \tt kniehl@mail.desy.de},
{\normalsize \tt kramer@mail.desy.de}, 
{\normalsize \tt schien@mail.desy.de}
\vspace{4mm} \\ 
and
\vspace{4mm} \\ 
{\bf H.~Spiesberger}
\vspace{2mm} \\
{\em \normalsize Institut f\"ur Physik, Johannes-Gutenberg-Universit\"at,}
\\ 
{\em \normalsize Staudinger Weg 7, D-55099 Mainz, Germany} 
\\
{\em \normalsize E-mail:} 
{\normalsize \tt hspiesb@thep.physik.uni-mainz.de}
}}

\date{}
\maketitle
\begin{abstract}
\medskip
\noindent
We present a detailed discussion of the collinear subtraction terms
needed to establish a massive variable-flavour-number scheme for the
one-particle inclusive production of heavy quarks in hadronic
collisions.  The subtraction terms are computed by convoluting
appropriate partonic cross sections with perturbative parton
distribution and fragmentation functions relying on the method of mass
factorization.  We find (with one minor exception) complete agreement
with the subtraction terms obtained in a previous publication by
comparing the zero-mass limit of a fixed-order calculation with the
genuine massles results in the $\msbar$ scheme.  This presentation will
be useful for extending the massive variable-flavour-number scheme to
other processes.
\end{abstract}


\clearpage


\section{Introduction}

Heavy-quark production in highly energetic $e^+ e^-$, $\gamma \gamma$,
$\gamma p$, $e p$ and $p \pbar$ collisions has attracted much interest
in the past few years, both experimentally and theoretically.  Heavy
quarks are those with masses $m \gg \Lambda_{\rm QCD}$ so that
$\alpha_s(m) \ll 1$, where $\alpha_s(\mu_R)$ is the strong-coupling
constant at renormalization scale $\mu_R$.  According to this
definition, the charm, bottom and top quarks ($c, b, t$) are heavy
whereas the up, down and strange quarks ($u, d, s$) are light.  Since
$\alpha_s(m) \ll 1$, heavy-quark production is a calculable process in
perturbative QCD. The heavy-quark mass acts as a cutoff for initial- and
final-state collinear singularities and sets the scale for the
perturbative expansion in $\alpha_s$.

On this basis, most of the next-to-leading-order (NLO) QCD calculations
have been performed in the past
\cite{Nason:1987xz,Nason:1989zy,Beenakker:1988bq,Beenakker:1990ma}.
Corresponding results are reliable when $m$ is the only large scale, as
for example in calculations of the total cross section or if any
additional scale, for example the transverse momentum $p_T$ of the
produced heavy quark in $\gamma \gamma$, $\gamma p$ and $p \pbar$
reactions or the lepton momentum transfer $Q$ in deep-inelastic $e p$
scattering (DIS), is not much larger than the mass $m$.  However, when
$p_T$ (or $Q$) is much larger than the mass, large logarithms $\ln
(p_T^2/m^2)$ or $\ln (Q^2/m^2)$ arise to all orders, so that fixed-order
perturbation theory is no longer valid.  As is well known, these
logarithms can be resummed and, this way, the perturbation series can be
improved.

The isolation and resummation of large logarithms is similar to the
conventional massless parton model approach, where initial- or
final-state collinear singularities are absorbed into parton
distribution functions (PDF) of the incoming hadrons or photons and into
fragmentation functions (FF) for the produced light hadrons (or
photons), respectively.  Therefore, this approach is usually referred to
as {\it zero-mass variable-flavour-number scheme} (ZM-VFNS).  The notion
``variable flavour number'' is used since, in the parton model, the
number of active quark flavours is increased by one unit, $n_f \to n_f +
1$, when the factorization scale crosses certain transition scales
(which are usually taken to be of the order of the heavy-quark
mass)~\footnote{For a detailed discussion see the appendix in Ref.\ 
  \protect\cite{Amundson:2000vg} and references given there.}.  In
contrast, the fixed-order treatment, where $m$ is kept as a large scale,
is called the {\it fixed-flavour-number scheme} (FFNS), since the number
of flavours in the initial state is fixed to $n_f =3$ $(4)$ for charm
(bottom) production.  One can combine cross sections calculated in the
FFNS after certain modifications with heavy-quark FFs and PDFs which
contain the resummed large logarithms.  This prescription is called the
{\it massive} or {\it general-mass} VFNS (GM-VFNS)~\footnote{For details
  see, e.g., Refs.\ \protect\cite{Tung:2001mv,Schienbein:2003et}.}.

One might expect that the partonic cross sections calculated in the FFNS
approach the corresponding ZM-VFNS cross sections in the limit $m \to 0$
if the collinear singular terms proportional to $\ln (m^2/s)$ are
subtracted, i.e., the subtracted FFNS cross sections differ from the
ZM-VFNS cross sections only by terms $\sim m^2/p_T^2$.  If this was
true, the FFNS and ZM-VFNS results for the cross sections would approach
each other for $p_T^2 \gg m^2$.  This expectation, however, is not true,
as was first demonstrated by Mele and Nason \cite{Mele:1991cw} for
inclusive heavy-quark production in $e^+e^-$ annihilation at NLO ($e^+
e^- \to Q \Qbar g$, where $Q$ is the heavy quark).  They found that the
limit $m \to 0$ of the cross section for $e^+ e^- \to Q \Qbar g$ and the
cross section calculated with $m=0$ from the beginning (in the $\msbar$
scheme) differ by finite terms of $\Ord(\alpha_s)$.  The reason for the
occurrence of these finite terms is the different definition of the
collinear singular terms in the two approaches.  In the ZM-VFNS
calculation, the heavy-quark mass is set to zero from the beginning and
the collinearly divergent terms are defined with the help of dimensional
regularization.  This fixes the finite terms in a specific way (in a
given factorization scheme), and their form is inherent to the chosen
regularization procedure.  If, on the other hand, one starts with $m \ne
0$ and performs the limit $m \to 0$ afterwards, the finite terms can be
different.  In Ref.\ \cite{Mele:1991cw}, it was shown that these
additional finite terms emerging in the limit $m \to 0$ out of the
theory with $m \ne 0$ can be generated in the theory for $m=0$ with
$\msbar$ factorization by convoluting this cross section with a partonic
fragmentation function $d_{Q\to Q}(x,\mu)$ for the transition from
massless to massive heavy quarks $Q$ (the explicit form of $d_{Q\to
  Q}(x,\mu)$ will be given later).

If this interpretation of the finite terms in the theory with $m\ne 0$
as partonic FF is generally true, then $d_{Q\to Q}(x,\mu)$ should be
process independent and could be used in any other heavy-quark
production process.  The universality of the partonic FF has been
confirmed by performing the same calculation as in Ref.\ 
\cite{Mele:1991cw} for the process $\gamma^\star Q \to Q g$
\cite{Kretzer:1998ju,Kretzer:1998nt}, where $\gamma^\star$ is a
space-like virtual photon, $\gamma \gamma \to Q \Qbar g$
\cite{Kramer:2001gd} and $g g \to Q \Qbar g$ \cite{Kniehl:2004fy} and
showing that the finite terms are obtained from a convolution of the
corresponding LO cross sections with $d_{Q\to Q}(x,\mu)$. The
process-independence of $d_{Q \to Q}(x,\mu)$ was established on more
general grounds in Ref.\ \cite{Cacciari:2001}. Moreover,
process-independent derivations of the partonic FFs have been performed
by Ma \cite{Ma:1997yq} and recently by Melnikov and Mitov
\cite{Melnikov:2004bm,Mitov:2004du}, who have computed the partonic FFs
to $\Ord(\alpha_s^2)$.

The fact that the theory with $m \ne 0$ and the ZM-VFNS are related by
the convolution of the ZM-VFNS cross section with partonic FFs has been
used in several ways.  In Ref.\ \cite{Mele:1991cw}, $d_{Q\to
  Q}(x,\mu_0)$ was used as the initial condition, at $\mu = \mu_0 =
\Ord(m)$, for the calculation of $d_{Q\to Q}(x,\mu)$ at an arbitrary
scale $\mu$ with the standard evolution equation.  Later, Cacciari and
Greco calculated with the same procedure the cross section for
heavy-quark production in $p \pbar$ and $p p$ collisions from the
ZM-VFNS cross section supplemented with evolved $d_{Q\to Q}(x,\mu)$ as a
function of $p_T$ \cite{Cacciari:1993mq}.  Partonic FFs used together
with a zero-mass hard-scattering cross section have subsequently been
applied also to heavy-quark production in $\gamma \gamma$
\cite{Cacciari:1995ej} and $\gamma p$
\cite{Cacciari:1996fs,Cacciari:1997du} processes.  In Ref.\ 
\cite{Cacciari:1997du}, the approach was generalized to the reaction
$\gamma + p \to D^\star + X$.  The transition $c \to D^\star$ was
described by a FF containing besides a non-perturbative contribution the
purely perturbative partonic FF. The non-perturbative part was described
by a function containing two parameters which were fixed by comparison
to experimental data for $e^+ + e^- \to D^\star + X$.  In Refs.\ 
\cite{Cacciari:1993mq,Cacciari:1995ej,Cacciari:1996fs,Cacciari:1997du},
the perturbative FF approach was motivated by the requirement to match
the ZM-VFNS as close as possible to the $m \ne 0$ theory.  This could be
achieved since at small $p_T = \Ord(m)$ the evolution of $d_{Q\to
  Q}(x,\mu_0)$ was not yet effective and, therefore, $d_{Q\to Q}$ was
just taking account of the difference of the two theories.  However,
terms proportional to $m^2/p_T^2$ are not included in this way.

The so-called FONLL approach
\cite{Cacciari:1998it,Cacciari:2001td,Cacciari:2002pa,Frixione:2002zv,
  Cacciari:2003zu} is an attempt to repair this deficiency. There, the
ZM-VFNS with perturbative FFs together with a non-perturbative component
was combined with the FFNS with $n_f = 3$ $(4)$ for charm (bottom)
production, introducing a $p_T$ dependent suppression factor by hand.
In addition, $m^2/p_T^2$ terms have been included in extensions of the
ACOT scheme \cite{Aivazis:1994kh,Aivazis:1994pi} to one-particle
inclusive production of $D$ mesons in charged-current and
neutral-current DIS \cite{Kretzer:1998nt,Kretzer:1997pd}.  In Ref.\ 
\cite{Olness:1997yc}, the ACOT scheme has been applied to one-particle
inclusive heavy-quark production in $p \pbar$ collisions.

Instead of incorporating the finite terms $d_{Q\to Q}(x,\mu)$ into the
initial conditions of the perturbative FFs at $\mu = \mu_0 = \Ord(m)$,
one can take this difference also into account by switching to a new
factorization scheme, which we call the massive factorization scheme.
In this scheme, starting from the ZM-VFNS, one adjusts the factorization
of the final-state collinear singularities associated with the massive
quarks in such a way that it matches the massive calculation in the
limit $m \to 0$.  Of course, the hard-scattering cross sections of any
other process for inclusive $D^\star$ production must be transformed to
the new scheme as well.  This is particularly important for the reaction
$e^+ + e^- \to D^\star + X$ from which the information on the
non-perturbative FF for $c \to D^\star$ is obtained by comparison to
experimental data.  So far, calculations in this massive factorization
scheme were performed for $\gamma + p \to D^\star + X$ in Ref.\ 
\cite{Binnewies:1997gz}, where also fits of the new FFs for $c \to
D^\star$ have been presented (for a comparison of FFs in the massive and
the ${\rm \overline{MS}}$ scheme, see Ref.\ \cite{Kramer:2000xx}).

The simplest way to connect the truly massless cross sections in the
$\msbar$ scheme with the massive cross sections is to subtract the
finite pieces $d_{Q\to Q}(x,\mu_0)$ from the massive theory. In this
way, one can incorporate also the $m^2/p_T^2$ terms, as given in the
massive theory, with the advantage that the massive theory approaches
the ZM-VFNS theory in the limit $p_T \to \infty$ or $m \to 0$. In
addition, by including also the terms proportional to $\ln m^2$
contained in $d_{Q\to Q}(x,\mu)$ one can obtain not only the finite
subtraction terms but also the terms needed for a transition to a new
factorization scale.  This approach has been applied to $\gamma + \gamma
\to D^\star + X$ \cite{Kramer:2001gd,Kramer:2003cw}, to $\gamma + p \to
D^\star + X$ \cite{Kramer:2003jw} and to $p + \pbar \to D^\star + X$
\cite{Kniehl:2004fy}.  In particular in Ref.\ \cite{Kniehl:2004fy}, we
obtained the finite subtraction terms by comparing the cross sections of
the massive theory, worked out by Bojak and Stratmann
\cite{Bojak:2001fx,Bojak-PhD}, in the limit $m \to 0$ with the cross
sections in the genuine massless theory in the $\msbar$ factorization
scheme as deduced by Aversa et al.\ \cite{Aversa:1988vb} in a form which
is equivalent to the convolution of the massless cross section with
$d_{Q\to Q}(x,\mu)$.

We are going to present details of this quite involved calculation in
this paper. The purpose is, on the one hand, to exactly demonstrate that
all the subtraction terms are generated by the convolution with partonic
FFs, at NLO just with $d_{Q\to Q}(x,\mu)$.  On the other hand, we hope
that the detailed presentation will show how the calculation carries
over to other processes $a + b \to D^\star + X$.  Since heavy-quark
production in hadron-hadron collisions is the most complex case, we
shall concentrate on this particular process.  Some results will also be
directly relevant for heavy-quark production in $\gamma \gamma$ and
$\gamma p$ processes.

The outline of this paper is as follows.  In Sec.\ \ref{sec:hadro}, we
consider heavy-quark production in hadronic collisions, introduce the
notation and review the derivation of the subtraction terms in Ref.\ 
\cite{Kniehl:2004fy}.  In Sec.\ \ref{sec:convolution}, we present the
convolution formulas, from which, in Sec.\ \ref{sec:subtraction}, the
various subtraction terms are calculated and compared with the results
in Ref.\ \cite{Kniehl:2004fy}.  Section \ref{sec:summary} contains a
summary and some concluding remarks.  The subprocess cross sections
needed in the convolutions have been collected in App.\ 
\ref{app:subprocesses} for convenience.


\section{Hadroproduction of heavy quarks}
\label{sec:hadro}

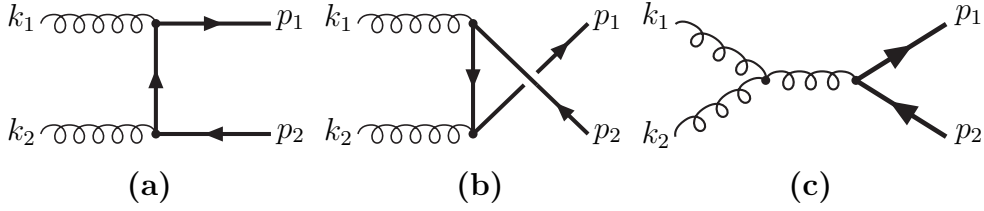
\begin{figure}[t]
\begin{center}
\setlength{\unitlength}{1pt}
\SetScale{0.85}
\begin{picture}(362,130)(0,0)
\put(5,20){
  \begin{picture}(102,63)
    \SetWidth{0.7}
    \Gluon(0,56)(51,56){4}{5}
    \Gluon(0,7)(51,7){4}{5}
    \Vertex(51,56){2.0}
    \Vertex(51,7){2.0}
    \SetWidth{1.8}
    \ArrowLine(102,7)(51,7)
    \ArrowLine(51,7)(51,56)
    \ArrowLine(51,56)(102,56)
    \Text(2,50)[r]{$k_1~$}
    \Text(2,7)[r]{$k_2~$}
    \Text(82,49)[l]{$~\;p_1$}
    \Text(82,6)[l]{$~\;p_2$}
  \end{picture}
}
\Text(50,0)[b]{\bf (a)}
\SetScale{0.85}
\put(125,20) {
  \begin{picture}(102,63)
    \SetWidth{0.7}
    \Gluon(0,56)(51,56){4}{5}
    \Gluon(0,7)(51,7){4}{5}
    \Vertex(51,56){2.0}
    \Vertex(51,7){2.0}
    \SetWidth{1.8}
    \Line(51,7)(73.67,28.78)
    \ArrowLine(79.33,34.22)(102,56)
    \ArrowLine(51,56)(51,7)
    \ArrowLine(102,7)(79.33,28.78)
    \Line(79.33,28.78)(51,56)
    \Text(2,50)[r]{$k_1~$}
    \Text(2,7)[r]{$k_2~$}
    \Text(82,49)[l]{$~\;p_1$}
    \Text(82,6)[l]{$~\;p_2$}
  \end{picture}
}
\Text(175,0)[b]{\bf (b)}
\SetScale{1.7}
\put(245,25){
  \begin{picture}(60,25)
    \SetWidth{0.47}
    \Gluon(0,25)(20,12.5){2}{3}
    \Gluon(0,0)(20,12.5){2}{3}
    \Gluon(20,12.5)(40,12.5){2}{3}
    \Vertex(20,12.5){1.0}
    \Vertex(40,12.5){1.0}
    \SetWidth{1.06}
    \ArrowLine(60,0)(40,12.5)
    \ArrowLine(40,12.5)(60,25)
    \Text(2,46)[r]{$k_1~$}
    \Text(2,1)[r]{$k_2~$}
    \Text(102,46)[l]{$~p_1$}
    \Text(102,1)[l]{$~p_2$}
  \end{picture}
}
\Text(300,0)[b]{\bf (c)}
\end{picture}
\end{center}
\caption{Feynman diagrams for the LO gluon-gluon fusion process $g + 
  g\rightarrow Q + \overline{Q}$.} 
\label{fig:gglo}
\end{figure}
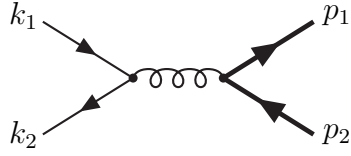
\begin{figure}[t!]
\begin{center}
\SetScale{1.7}
\setlength{\unitlength}{1pt}
\begin{picture}(102,60)
  \SetWidth{0.47}
  \ArrowLine(0,25)(20,12.5)
  \ArrowLine(20,12.5)(0,0)
  \Gluon(20,12.5)(40,12.5){2}{3}
  \Vertex(20,12.5){1.0}
  \Vertex(40,12.5){1.0}
  \SetWidth{1.06}
  \ArrowLine(60,0)(40,12.5)
  \ArrowLine(40,12.5)(60,25)
  \Text(2,45)[r]{$k_1~$}
  \Text(2,0)[r]{$k_2~$}
  \Text(102,45)[l]{$~p_1$}
  \Text(102,0)[l]{$~p_2$}
\end{picture}
\end{center}
\caption{The LO quark-antiquark annihilation process $q + \overline{q}
  \rightarrow Q + \overline{Q}$.}  
\label{fig:qqlo}
\end{figure}

In the FFNS, the following partonic subprocesses contribute to $p +
\pbar \to H + X$ in leading order (LO) and NLO, where $H = D,$
$D^\star$, $B\ldots$ is a heavy meson:
\begin{itemize}
\item[1.] $g(k_1) + g(k_2) \to Q(p_1) + \Qbar(p_2) + [g(p_3)]$, where
  $Q=c,b$ denotes a heavy quark. The LO Feynman diagrams are shown in
  Fig.\ \ref{fig:gglo}.
\item[2.] $q(k_1) + \qbar(k_2) \to Q(p_1) + \Qbar(p_2) + [g(p_3)]$. In
  LO, there is one Feynman diagram, which is shown in Fig.\ 
  \ref{fig:qqlo}.
\item[3.] $g(k_1) + q(k_2) \to Q(p_1) + \Qbar(p_2) + q(p_3)$ and $g(k_1)
  + \qbar(k_2) \to Q(p_1) + \Qbar(p_2) + \qbar(p_3)$ contribute at NLO.
  The Feynman diagrams for these processes, as well as those for the NLO
  contributions of $gg \to Q\Qbar g$ and $q\qbar \to Q\Qbar g$, can be
  found in App.\ \ref{app:feynman}.
\end{itemize}

Our aim is to calculate differential cross sections with an observed
heavy quark $Q$ of momentum $p_1$. Therefore we define the following
invariants
\begin{eqnarray}
 s &=& (k_1+k_2)^2\, , \nonumber \\
 t_1&=& t-m^2 = (k_1-p_1)^2 - m^2\, , \nonumber \\
 u_1 &=& u-m^2 = (k_2-p_1)^2 - m^2\, , 
\end{eqnarray}
and
\begin{equation}
 s_2 = S_2 - m^2 = (k_1+k_2-p_1)^2 - m^2 = s+t_1+u_1
\end{equation}
with $s+t_1+u_1=0$ at LO, where $p_3=0$.  As usual, we introduce the
dimensionless variables $v$ and $w$ by
\begin{equation}
v = 1 + \frac{t_1}{s}, ~~~~ w = - \frac{u_1}{s+t_1}, 
\end{equation}
so that
\begin{equation}
t_1 = -s (1-v), ~~~~ u_1 = - s v w\ .
\end{equation}
In LO, we have $w=1$.

In a recent publication \cite{Kniehl:2004fy}, we have presented a NLO
calculation for the inclusive production of $D^\star$ mesons in $p
\pbar$ collisions including heavy-quark mass effects in the
hard-scattering cross sections.  The following procedure has been
adopted \cite{Kramer:2001gd,Kramer:2003cw} (see also Refs.\ 
\cite{Schienbein:2003et,Schienbein:2004ah}):
\begin{itemize}
\item[(i)] We have derived the zero-mass limit of the cross sections in
  the massive FFNS with $n_f = 3$ \cite{Bojak:2001fx,Bojak-PhD} for the
  partonic subprocesses listed above only keeping $m$ as a regulator in
  logarithms $\ln \left(m^2/s\right)$.  Special care was required in
  order to recover the distributions $\delta(1-w)$,
  $\left(1/(1-w)\right)_+$ and $\left(\ln(1-w)/(1-w)\right)_+$ occurring
  in the massless $\msbar$ calculation, see, e.g., Eq.\ (12) in Ref.\ 
  \cite{Kniehl:2004fy}. The result, generically denoted $\lim_{m \to 0}
  \der \tilde{\sigma}(m)$, contains mass singular logarithms $\ln
  (m^2)$, but collinear singularities associated with light quarks and
  gluons are already subtracted in $\der \tilde{\sigma}(m)$.
\item[(ii)] Then we have compared the massless limit with the
  corresponding 
  hard-scattering cross sections
  in the genuine massless $\msbar$ calculation in order to identify
  appropriate subtraction terms.  Generically, one can write
  \begin{equation}
  \der \sigma^{\sub} = \lim_{m \to 0} \der \tilde{\sigma}(m)
  - \der \hat{\sigma}_{\msbar}\ ,
  \label{eq:sigmasub}
  \end{equation}
  where $\der \hat{\sigma}_{\msbar}$ is a hard-scattering cross section
  in the genuine $\msbar$ calculation.
\item[(iii)] The desired massive hard-scattering cross sections have
  then been constructed by removing the subtraction terms from the
  massive cross sections in the fixed-order theory:
  \begin{equation}
  \der \hat{\sigma}(m) = \der \tilde{\sigma}(m) - \der \sigma^{\sub} \, .
  \label{eq:hard1}
  \end{equation}
  By this procedure, the collinear logarithms $\ln (m^2/s)$ along with
  finite terms which are independent of the heavy-quark mass are
  subtracted from $\der \tilde{\sigma}(m)$.  On the other hand, all
  finite mass terms of the form $(m^2/p_T^2)^n$ (with an integer $n$)
  are kept in the hard-scattering cross sections.
\item[(iv)] Contributions with charm quarks in the {\em initial state}
  have been included in the massless approach.  It can be shown that
  neglecting the corresponding heavy-quark mass terms corresponds to a
  convenient choice of scheme (S-ACOT scheme \cite{Kramer:2000hn}) which
  does not imply any loss of precision.  In fact, the error which is
  made is of the same order as the error of the factorization formula,
  as has been discussed in the context of heavy-quark production in deep
  inelastic scattering \cite{Kramer:2000hn,Collins:1998rz}.  Obviously,
  this rule is of great practical importance since the existing massless
  results for the hard-scattering cross sections \cite{Aversa:1988vb}
  can simply be used, whereas their massive analogues are unknown and
  would require a dedicated calculation of these processes.\footnote{For
    deep inelastic scattering, massive-quark-initiated coefficients have
    been obtained in Refs.\ \cite{Kretzer:1998ju,Kretzer:1998nt}; the
    results for this simpler case are already quite involved.}
\end{itemize}

Note that also the FONLL calculation in Ref.\ \cite{Cacciari:1998it} has
been constructed with the help of the zero-mass limit of the fixed-order
calculation in Refs.\ \cite{Nason:1987xz,Nason:1989zy}.  On the other
hand, in the GM-VFNS of Ref.\ \cite{Olness:1997yc}, the collinear
subtractions have been obtained using the methods of mass factorization
in a massive regularization scheme. In this approach, the subtraction
terms are computed by convolutions of appropriate subprocesses with
universal partonic PDFs and FFs.  However, the discussion in Ref.\ 
\cite{Olness:1997yc} is rather generic without presenting many details.
It is the purpose of this paper to provide a detailed description of the
derivation of the collinear subtraction terms using the convolution
method and to compare with the results obtained in our previous
publication \cite{Kniehl:2004fy}.


\section{Mass factorization with massive quarks}
\label{sec:convolution}

The starting point in our approach is the basic factorization formula at
the partonic level:
\begin{eqnarray}
\der \tilde\sigma(a + b \to Q + X) = 
f_{a \to i}(x_1) \otimes
f_{b \to j}(x_2) \otimes
\der \hat{\sigma}(i + j \to k + X) \otimes
d_{k \to Q}(z) \, ,
\label{eq:fact}
\end{eqnarray}
where $\der \tilde\sigma$ denote partonic cross sections (with
singularities due to light-quark and gluon lines already subtracted via
conventional mass factorization \cite{Ellis:1979xx}) and $\der
\hat{\sigma}$ are IR-safe hard-scattering cross sections which are free
of logarithms of the heavy-quark mass. The indices $a$, $b$, and $i$,
$j$, $k$ denote partons, and a sum over double indices is implied here
and in the following.  All logarithms of the heavy-quark mass (i.e. the
mass singularities in the zero-mass limit) are contained in the partonic
distribution functions $f_{a \to i}$ and in the partonic fragmentation
functions $d_{k \to Q}$.  The convolution $\otimes$ denotes the usual
convolution integral and will be specified below in Eqs.\ 
\eqref{eq:subx1}, \eqref{eq:subx2} and \eqref{eq:subz}.  Equation
\eqref{eq:fact} reflects the fact that the partonic cross sections $\der
\tilde\sigma$ can be factorized into process-dependent IR-safe
hard-scattering cross sections $\der \hat{\sigma}$, which are
well-behaved and finite in the limit $m \to 0$, and universal
(process-independent) partonic PDFs $f_{a \to i}$ and partonic FFs $d_{k
  \to Q}$.

Equation \eqref{eq:fact} can be expanded in powers of $\alpha_s$ with
the help of
\begin{eqnarray}
f_{a \to i}(x_1) 
&=& \delta_{ia} \delta(1-x_1) + f_{a \to i}^{(1)}  + f_{a \to i}^{(2)}
+ \ldots \, ,
\nonumber\\
f_{b \to j}(x_2) 
&=& \delta_{jb} \delta(1-x_2) + f_{b \to j}^{(1)}  + f_{b \to j}^{(2)}
+ \ldots \, ,
\nonumber\\
d_{k \to Q}(z) 
&=& \delta_{kQ} \delta(1-z) + d_{k \to Q}^{(1)}  + d_{k \to Q}^{(2)}
+ \ldots \, ,
\\
\der \hat{\sigma} &=& \der \hat{\sigma}^{(0)} + \der \hat{\sigma}^{(1)} 
+ \der \hat{\sigma}^{(2)} + \ldots \, ,
\nonumber\\
\der \tilde{\sigma} &=& \der \tilde{\sigma}^{(0)} + \der \tilde{\sigma}^{(1)} 
+ \der \tilde{\sigma}^{(2)} + \ldots \, .
\nonumber
\label{eq:expansion}
\end{eqnarray}
For the partonic PDFs and FFs, the superscript denotes the order of
$\alpha_s$.  For the cross sections, it indicates the relative order in
$\alpha_s$ with respect to the Born cross sections.  The expansion of
Eq.\ \eqref{eq:fact} can be used to determine order by order the
relation between the hard-scattering and partonic cross sections. Up to
NLO, one finds
\begin{eqnarray}
\der \hat{\sigma}^{(0)}(a + b \to Q + X) &=& 
\der \tilde{\sigma}^{(0)} (a + b \to Q + X) = 
\der {\sigma}^{(0)} (a + b \to Q + X) \, ,
\\
\der \hat{\sigma}^{(1)}(a + b \to Q + X) 
&=& 
\der \tilde{\sigma}^{(1)}(a + b \to Q + X)
\nonumber \\ & &
- f_{a \to i}^{(1)}(x_1) \otimes \der \sigma^{(0)}(i + b \to Q + X) 
\nonumber \\ & &
- f_{b \to j}^{(1)}(x_2) \otimes \der \sigma^{(0)}(a + j \to Q + X) 
\label{eq:hard2}
\\ & &
- \der \sigma^{(0)}(a + b \to k + X) \otimes d_{k \to Q}^{(1)}(z) \, .
\nonumber
\end{eqnarray}
The three convolutions in Eq.\ \eqref{eq:hard2} can be identified with
the subtraction term $\der \sigma^{\sub}$ in Eq.\ \eqref{eq:hard1}.

The factorization in Eq.\ \eqref{eq:fact} has to be defined at a
definite energy or momentum scale which enters as an argument into the
PDFs, FFs and $\der \sigmahat$.  We denote the factorization scales by
$\mui$ for initial-state factorization (entering the PDFs) and by $\muf$
for final-state factorization (entering the FFs). The renormalization
scale will be called $\mur$.


\subsection{Partonic parton distribution and fragmentation functions}
\label{sec:transition}

The functions $f^{(1)}_{i \to j}$ for the initial state are given in the
$\msbar$ scheme\footnote{Note that it is assumed that the $\msbar$
  scheme is defined in the {\em conventional} way where photons and
  gluons have $d-2$ degrees of freedom (where $d$ is the number of
  space-time dimensions).  Furthermore, subtractions $f_{ij} \otimes
  \der \sigma^{(0)}$ are performed with subprocess cross sections
  calculated in $d$ dimensions.}, keeping the heavy-quark mass as a
regulator for the collinear divergences, by
\begin{eqnarray}
\fcg(x,\mur,\mui) 
& = &
\displaystyle
\frac{\alpha_s(\mur)}{2 \pi} P^{(0)}_{g \to q}(x) 
\ln \frac{\muis}{m^2}\ ,
\nonumber\\
\fcc(x,\mur,\mui) 
& = &
\displaystyle
\frac{\alpha_s(\mur)}{2 \pi} 
C_F \left[\frac{1+x^2}{1-x} 
\left(\ln \frac{\muis}{m^2}- 2 \ln(1-x)-1\right)\right]_+ \ ,
\label{eq:initial}
\\
\fgg(x,\mur,\mui) 
& = &
\displaystyle
-\frac{\alpha_s(\mur)}{2 \pi}
\frac{2}{3} T_f
\ln \frac{\muis}{m^2} \delta(1-x)\ ,
\nonumber
\end{eqnarray}
where $P^{(0)}_{g \to q}(x)= \tfrac{1}{2}[x^2+(1-x)^2]$ and
$P^{(0)}_{q\to q}(x)=C_F [(1+x^2)/(1-x)]_+$ (appearing in $\fcc$) are
the conventional (space-like) one-loop splitting functions and
$T_f=1/2$.  The function $\fcc(x,\mur,\mui)$ will not be used in the
following, since heavy quarks in the initial state are treated as
massless quarks as explained in Sec.\ \ref{sec:hadro}.  It would be
present in cases where massive heavy quarks $Q$ appear in the initial
state as for example in Refs.\ 
\cite{Kretzer:1998ju,Kretzer:1998nt,Aivazis:1994kh,Aivazis:1994pi}.

The functions $d^{(1)}_{i\to j}$ for the final state read
\cite{Mele:1991cw,Kretzer:1998ju,Kretzer:1998nt,Ma:1997yq}
\begin{eqnarray}
\dcg(z,\mur,\muf) 
& = &
\frac{\alpha_s(\mur)}{2 \pi} P^{(0)}_{g \to q}(z) 
\ln \frac{\mufs}{m^2}\ ,
\nonumber\\
\dcc(z,\mur,\muf) 
& = &
\frac{\alpha_s(\mur)}{2 \pi} 
C_F \left[\frac{1+z^2}{1-z} 
\left(\ln \frac{\mufs}{m^2}- 2 \ln(1-z)-1\right)\right]_+ \ .
\label{eq:final}
\end{eqnarray}
Generally, the splitting functions entering the partonic FFs are
time-like splitting functions which are, however, identical to the
space-like splitting functions at the one-loop level.  It should be
noted that the function $\fcc(x,\mur,\mui)$ in Eq.\ \eqref{eq:initial}
is of the same form as $\dcc(x,\mur,\muf)$ at $\Ord(\alpha_s^1)$
\cite{Kretzer:1998ju,Kretzer:1998nt}. This will not be true at higher
orders since the NLO space- and time-like splitting functions $P_{q \to
  q}^{(1)}(x)$ are different.  All the other partonic PDFs and FFs not
listed here are zero at $\Ord(\alpha_s^1)$.  Furthermore, analogous
results for processes involving photon splittings can be found by
obvious replacements ($g \to \gamma$, $\alpha_s \to \alpha$ and
appropriate modifications of colour factors) in Eqs.\ \eqref{eq:initial}
and \eqref{eq:final}.

The partonic PDFs and FFs are known to order $\Ord(\alpha_s^2)$.  They
would be needed, together with the three-loop beta function of QCD, for
computing subtraction terms at next-to-NLO (NNLO).  For the initial
state, the partonic PDFs at order $\Ord(\alpha_s^2)$ can be found in
Ref.\ \cite{Buza:1998wv} (with the exception of $f_{Q \to Q}^{(2)}(x)$,
which is unknown).  Recently, also the $\Ord(\alpha_s^2)$ contributions
to the perturbative FFs have been derived
\cite{Melnikov:2004bm,Mitov:2004du}.  It should be noted that, at
$\Ord(\alpha_s^2)$, all of the perturbative PDFs and FFs are
non-vanishing at $\mui = m$ and $\muf = m$, respectively.  In fact, the
parts proportional to logarithms of the factorization scale follow from
the evolution equations, so that the new information obtained from the
$\Ord(\alpha_s^2)$ calculation is contained in the non-vanishing pieces
at $\mui, \muf = m$.

In Sec.\ \ref{sec:subtraction}, we will need the distribution
$\dcc(\zb)$ with $\zb = 1-v+v w$ as a distribution in the kinematic
variables $v$ and $w$. This form of $\dcc(\zb)$ can be written as:
\begin{equation}
\dcc(\zb) = A(v)\ \delta(1-w) + B(v)\  \frac{1}{(1-w)}_+
+ C(v)  \left(\frac{\ln(1-w)}{(1-w)}\right)_+ + D(v,w) \, ,
\label{eq:dcc}
\end{equation}
with
\begin{eqnarray*}
A(v) &=& \CF \frac{\alpha_s(\mur)}{2 \pi} \frac{1}{2 v}
\left[\Lfin (3 + 4 \ln v) + 4(1-\ln v - \ln^2 v) \right]\, ,
\\
B(v) &=& \CF \frac{\alpha_s(\mur)}{2 \pi} \frac{2}{v}
\left[\Lfin - 1 - 2 \ln v \right]\, ,
\\
C(v) &=& -\CF \frac{\alpha_s(\mur)}{2 \pi} \frac{4}{v}\, ,
\\
D(v,w) &=& -\CF \frac{\alpha_s(\mur)}{2 \pi}\left[2- v(1-w)\right]
\left[
\Lfin - 1 - 2 \ln v - 2 \ln (1-w) 
\right]\, .
\end{eqnarray*}


\subsection{Subtraction terms at NLO}
\label{sec:subatnlo}

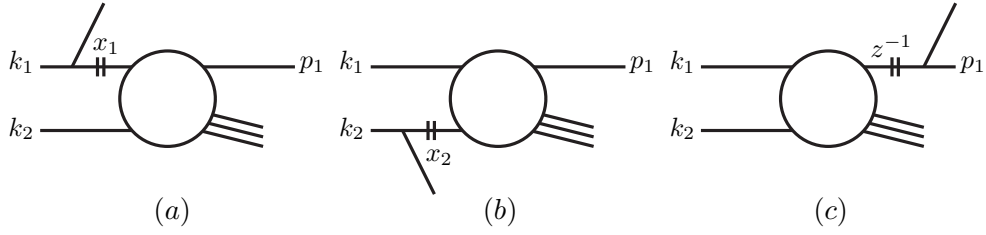
\begin{figure}[t]
\begin{center}
\setlength{\unitlength}{1pt}
\begin{picture}(360,90)(-30,0)
\SetScale{1.2}
\SetWidth{1.0}
\put(-30,0){
  \begin{picture}(100,80)(0,0)
  \Line(10,50)(40,50)
  \Line(10,30)(40,30)
  \Line(60,50)(90,50)
  \Line(20,50)(30,70)
  \Line(28,53)(28,47)
  \Line(30,53)(30,47)
  \Line(60,30)(80,25)
  \Line(60,33)(80,28)
  \Line(60,36)(80,31)
  \IfColor{\COval(50,40)(15,15)(0){Black}{White}}{\GOval(50,40)(15,15)(0){0.5}}
  \Text(37,68)[c]{\footnotesize $x_1$}
  \Text(5,62)[c]{\footnotesize $k_1$}
  \Text(5,37)[c]{\footnotesize $k_2$}
  \Text(115,61)[c]{\footnotesize $p_1$}
  \end{picture}
}
\Text(36,0)[b]{\small\bf $(a)$}
\put(95,0){
  \begin{picture}(100,80)(0,0)
  \Line(10,50)(40,50)
  \Line(10,30)(40,30)
  \Line(60,50)(90,50)
  \Line(20,30)(30,10)
  \Line(28,33)(28,27)
  \Line(30,33)(30,27)
  \Line(60,30)(80,25)
  \Line(60,33)(80,28)
  \Line(60,36)(80,31)
  \IfColor{\COval(50,40)(15,15)(0){Black}{White}}{\GOval(50,40)(15,15)(0){0.5}}
  \Text(38,26)[c]{\footnotesize $x_2$}
  \Text(5,62)[c]{\footnotesize $k_1$}
  \Text(5,37)[c]{\footnotesize $k_2$}
  \Text(115,61)[c]{\footnotesize $p_1$}
  \end{picture}
}
\Text(160,0)[b]{\small $(b)$}
\put(220,0){
  \begin{picture}(100,80)(0,0)
  \Line(10,50)(40,50)
  \Line(10,30)(40,30)
  \Line(60,50)(90,50)
  \Line(80,50)(90,70)
  \Line(72,53)(72,47)
  \Line(70,53)(70,47)
  \Line(60,30)(80,25)
  \Line(60,33)(80,28)
  \Line(60,36)(80,31)
  \IfColor{\COval(50,40)(15,15)(0){Black}{White}}{\GOval(50,40)(15,15)(0){0.5}}
  \Text(84,68)[c]{\footnotesize $z^{-1}$}
  \Text(5,62)[c]{\footnotesize $k_1$}
  \Text(5,37)[c]{\footnotesize $k_2$}
  \Text(115,61)[c]{\footnotesize $p_1$}
  \end{picture}
}
\Text(285,0)[b]{\small $(c)$}
\end{picture}
\end{center}
\caption{Sketch of kinematics of mass factorization for (a) upper
  incoming line (b) lower incoming line and (c) outgoing line.}
\label{fig:fact}
\end{figure}

We distinguish mass factorization in the initial state and in the final
state.  For one-particle inclusive production, where one final-state
particle has a fixed momentum (above, we had chosen $p_1$), we have to
distinguish further two cases with initial-state singularities
corresponding to $t$- and $u$-channel scattering. A graphical
representation of the subtraction terms in form of cut diagrams for all
cases is shown in Fig.\ \ref{fig:fact}.  These diagrams can be found by
applying all possible cuts to internal lines of the Feynman diagrams
(see App.\ \ref{app:feynman}).  The cuts are allowed if the $2 \to 2$
subprocesses are kinematically possible and the $1 \to 2$ process
involves the splitting into a heavy-quark line. In an axial gauge, the
cut diagrams correspond to actual Feynman diagrams.

\subsubsection{Initial-state factorization}

In the first case with $u$-channel scattering (see Fig.\ 
\protect\ref{fig:fact}(a)), the collinear subtractions are given by
\begin{eqnarray}
\der \sigma^\sub(a + b \to Q + X) 
&=& \int_0^1 \der x_1\ f_{a\to i}^{(1)}(x_1,\mur,\mui)\
\der \sigmahat^{(0)}\left(i(x_1 k_1) + b(k_2) \to Q(p_1) + X\right)
\nonumber\\
&\equiv& f_{a\to i}^{(1)}(x_1) \otimes \der \sigmahat^{(0)}(i + b \to Q
+ X)\, .
\label{eq:subx1}
\end{eqnarray}
Here $a + b \to Q + X$ stands for the one-particle inclusive partonic
subprocesses ($g + g \to Q + X$, $q + \qbar \to Q + X$, $g + q \to Q +
X$, $g + \qbar \to Q + X$), $f_{a\to i}(x_1,\mur,\mui)$ describes the
collinear splitting of parton '$a$' into parton '$i$', and $i + b \to Q
+ X$ are the corresponding $2 \to 2$ subprocesses with momenta $x_1
k_1$, $k_2$ and $p_1$.  A sum over $i$ is implied, i.e., all possible
splittings and subprocesses have to be taken into account.

We define the following invariants for the subprocess:
\begin{eqnarray}
\shat &=& (x_1 k_1+k_2)^2 = x_1 s \, ,
\nonumber \\
\that_1&=& (x_1 k_1-p_1)^2 - m^2 = x_1 t_1\, ,
\\
\uhat_1 &=& (k_2-p_1)^2 - m^2  = u_1\, ,
\nonumber
\end{eqnarray}
and
\begin{eqnarray}
\vhat &=& 1 + \frac{\that_1}{\shat} = v\quad , \quad
\what = - \frac{\uhat_1}{\shat+\that_1} = \frac{w}{x_1} \, ,
\\
\that_1 &=& -\shat (1-\vhat)\quad , \quad 
\uhat_1 = - \shat \vhat \what = - \shat \vhat \, .
\end{eqnarray}
For the calculation of $\der^2 \sigma^\sub/(\der v \der w)$ in Eq.\ 
\eqref{eq:subx1}, it is convenient to write the subprocess cross section
as
\begin{eqnarray}
\frac{\der^2 \sigmahat^{(0)}}{\der v \der w}
& = &
J \frac{\der^2 \sigmahat^{(0)}}{\der \vhat \der \what} 
=
J \frac{\der \sigmahat^{(0)}}{\der \vhat} \delta(1-\what)
\label{eq:sub-xs}
\end{eqnarray}
with 
\begin{equation}
\delta(1-\what) = w \delta(x_1-\xb_1) \, , \quad \quad \xb_1 = w \, .
\end{equation}
The $\delta$-function imposes the $2 \to 2$ process kinematics $\shat +
\that_1 + \uhat_1 = 0$, i.e.\ $\what = 1$, and implies $\xb_1 = w$.  The
Jacobian reads
\begin{equation}
J = \frac{\partial (\vhat,\what)}{\partial (v,w)} = \frac{1}{x_1}\ .
\end{equation}
Combining these results we find
\begin{equation}
\frac{\der^2 \sigmahat^{(0)}}{\der v \der w} 
= 
\left. \frac{\der \sigmahat^{(0)}}{\der \vhat}
\right|_{\shat\to \xb_1 s,\; \vhat \to v} \delta(x_1-\xb_1) \, ,
\end{equation}
so that the subtraction terms for initial-state factorization of the
upper incoming line can be calculated as
\begin{eqnarray}
\frac{\der^2 \sigma^\sub}{\der v \der w}(a + b \to Q + X) 
& = &
f^{(1)}_{a\to i}(\xb_1,\mur,\mui) 
\left.
\frac{\der \sigmahat^{(0)}}{\der \vhat}(i + b \to Q + X) 
\right|_{\vhat \to v,\; \shat \to \xb_1 s} \, .
\label{eq:factx1}
\end{eqnarray}

In the second case with $t$-channel scattering (see Fig.\ 
\protect\ref{fig:fact}(b)), the collinear subtractions are given by
\begin{eqnarray}
\der \sigma^\sub(a + b \to Q + X) &=& \int_0^1 \der x_2\ 
f_{b\to j}^{(1)}(x_2,\mur,\mui)\
\der \sigmahat^{(0)}(a(k_1) + j(x_2 k_2) \to Q(p_1) + X)
\nonumber\\
&\equiv& f_{b\to j}^{(1)}(x_2) \otimes \der \sigmahat^{(0)}(a + j \to Q
+ X)\, .
\label{eq:subx2}
\end{eqnarray}
The invariants for the subprocess are now given by
\begin{eqnarray}
\shat &=& (k_1+ x_2 k_2)^2 = x_2 s\, , 
\nonumber \\
\that_1&=& (k_1-p_1)^2 - m^2 = t_1\, ,
\\
\uhat_1 &=& (x_2 k_2-p_1)^2 - m^2  = x_2 u_1\, ,
\nonumber
\end{eqnarray}
and
\begin{eqnarray}
\vhat &=& \frac{x_2 - 1 + v}{x_2} \, , \quad \quad
\what = \frac{x_2 v w}{x_2 -1 + v} \, .
\end{eqnarray}
Again, we write the subprocess cross section as in Eq.\ 
\eqref{eq:sub-xs} with
\begin{equation}
\delta(1-\what) = \xb_2^2 \frac{v w}{1-v} \delta(x_2-\xb_2) 
\, , \quad \quad
\xb_2 = \frac{1-v}{1-v w}\, ,
\end{equation}
and
\begin{equation}
J = \frac{\partial (\vhat,\what)}{\partial (v,w)} 
= \frac{v}{x_2 -1 +v}\, .
\end{equation}
For the $2 \to 2$ subprocess kinematics, we have $\hat{w} = 1$, $x_2 =
\xb_2$, $\vhat = v w$ and $J=1 / (\xb_2 w)$.  Combining these results,
we find
\begin{equation}
\frac{\der^2 \sigmahat^{(0)}}{\der v \der w} 
= 
\frac{v}{1-v w}\
\left. 
\frac{\der \sigmahat^{(0)}}{\der \vhat}
\right|_{\shat\to \xb_2 s,\; \vhat \to v w} \delta(x_2-\xb_2) \, ,
\end{equation}
so that the subtraction terms for initial-state factorization of the
lower incoming line can be calculated as
\begin{eqnarray}
\label{eq:factx2}
\frac{\der^2 \sigma^\sub}{\der v \der w}(a + b \to Q + X) 
& = &
f^{(1)}_{b\to j}(\xb_2,\mur,\mui) \
\frac{v}{1-v w}\
\left.
\frac{\der \sigmahat^{(0)}}{\der \vhat}(a + j \to Q + X) 
\right|_{\vhat \to v w, \; \shat \to \xb_2 s} 
\end{eqnarray}


\subsubsection{Final-state factorization}

The case shown in Fig.\ \protect\ref{fig:fact}(c) corresponds to
factorization of singularities in the final state. Here the collinear
subtractions are given by
\begin{eqnarray}
\der \sigma^\sub(a + b \to Q + X) &=& \int_0^1 \der z\ 
\der \sigmahat^{(0)}\left(a(k_1) + b(k_2) \to k(z^{-1} p_1) + X\right)\ 
d_{k\to Q}^{(1)}(z,\mur,\muf)\
\nonumber\\
&\equiv& 
\der \sigmahat^{(0)}(a + b \to k + X) \otimes d_{k\to Q}^{(1)}(z) \, .
\label{eq:subz}
\end{eqnarray}
The invariants for the subprocess can be defined as follows
\begin{eqnarray}
\shat &=& (k_1+ k_2)^2 = s \, ,
\nonumber \\
\that_1&=& (k_1- z^{-1} p_1)^2 - m^2 = \frac{1}{z} t_1\, ,
\\
\uhat_1 &=& (k_2- z^{-1} p_1)^2 - m^2  = \frac{1}{z} u_1\, ,
\nonumber
\end{eqnarray}
and
\begin{eqnarray}
\vhat &=& \frac{z - 1 + v}{z} \, , \quad \quad
\what = \frac{v w}{z -1 + v} \, .
\end{eqnarray}
As before, we write the subprocess cross section as in Eq.\ 
\eqref{eq:sub-xs} with
\begin{equation}
\delta(1-\what) = v w \delta(z-\zb) \, , \quad \quad \zb = 1-v+vw \, ,
\end{equation}
and
\begin{equation}
J = \frac{\partial (\vhat,\what)}{\partial (v,w)} = 
\frac{1}{z}\frac{v}{z -1 +v}\, .
\end{equation}
From $\what = 1$ one finds $\zb = 1-v+v w$, $\vhat = v w / \zb$ and $J=1
/ (\zb w)$.  Combining these results, we find
\begin{equation}
\frac{\der^2 \sigmahat^{(0)}}{\der v \der w} 
= 
\frac{v}{\zb}\
\left. 
\frac{\der \sigmahat^{(0)}}{\der \vhat}
\right|_{\shat\to s,\; \vhat \to \frac{v w}{\zb}}
\delta(z-\zb) \, ,
\end{equation}
so that the subtraction terms for final-state factorization can be
calculated as
\begin{eqnarray}
\label{eq:factz}
\frac{\der^2 \sigma^\sub}{\der v \der w}(a + b \to Q + X) 
& = &
d_{k\to Q}^{(1)}(\zb,\mur,\muf) 
\frac{v}{\zb}
\left. 
\frac{\der \sigmahat^{(0)}}{\der \vhat}(a + b \to k + X) 
\right|_{\vhat \to v w/\zb, \; \shat \to s}
\end{eqnarray}


\subsection{Scheme dependence and implementation freedom}

Before we turn to a discussion of our results for the collinear
subtraction terms calculated according to Eqs.\ \eqref{eq:factx1},
\eqref{eq:factx2} and \eqref{eq:factz}, we add some additional remarks.

\begin{figure}[t]
\begin{center}
\setlength{\unitlength}{1pt}
\SetScale{0.85}
\begin{picture}(270,130)(0,0)
\put(5,20){
  \begin{picture}(115,63)
    \SetWidth{0.7}
    \Gluon(-30,53)(-2,53){3}{3}
    \Gluon(12,53)(38,53){3}{3}
    \Gluon(-30,7)(38,7){3}{8}
    \SetWidth{1.8}
    \ArrowLine(112,7)(62,7)
    \ArrowLine(62,53)(112,53)
    \SetWidth{0.7}
  \IfColor{\COval(50,30)(30,20)(0){Black}{White}}{\GOval(50,30)(30,20)(0){0.5}}
    \SetWidth{1.8}
  \IfColor{\COval(5,53)(7,7)(0){Black}{White}}{\GOval(5,53)(7,7)(0){0.5}}
    \SetWidth{0.7}
    \SetWidth{1.0}
    \Line(23,63)(23,43)
    \Line(26,63)(26,43)
    \Text(-25,48)[r]{$k_1~$}
    \Text(-25,7)[r]{$k_2~$}
    \Text(92,46)[l]{$~\;p_1$}
    \Text(92,6)[l]{$~\;p_2$}
  \end{picture}
}
\Text(50,0)[b]{\bf (a)}
\SetScale{0.85}
\put(170,20) {
  \begin{picture}(115,63)
    \SetWidth{0.7}
    \Gluon(-30,7)(-2,7){3}{3}
    \Gluon(12,7)(38,7){3}{3}
    \Gluon(-30,53)(38,53){3}{8}
    \SetWidth{1.8}
    \ArrowLine(112,7)(62,7)
    \ArrowLine(62,53)(112,53)
    \SetWidth{0.7}
  \IfColor{\COval(50,30)(30,20)(0){Black}{White}}{\GOval(50,30)(30,20)(0){0.5}}
    \SetWidth{1.8}
  \IfColor{\COval(5,7)(7,7)(0){Black}{White}}{\GOval(5,7)(7,7)(0){0.5}}
    \SetWidth{0.7}
    \SetWidth{1.0}
    \Line(23,17)(23,-3)
    \Line(26,17)(26,-3)
    \Text(-25,48)[r]{$k_1~$}
    \Text(-25,7)[r]{$k_2~$}
    \Text(92,46)[l]{$~\;p_1$}
    \Text(92,6)[l]{$~\;p_2$}
  \end{picture}
}
\Text(217,0)[b]{\bf (b)}
\end{picture}
\end{center}
\caption{Feynman diagrams representing (a)
  $\fgg(x_1) \otimes \der \sigmahat^{(0)}(g g \to Q \Qbar)$ and (b)
  $\fgg(x_2) \otimes \der \sigmahat^{(0)}(g g \to Q \Qbar)$.  The
  fermion loops on the external gluon lines are heavy-quark loops.  }
\label{fig:cutgg0}
\end{figure}
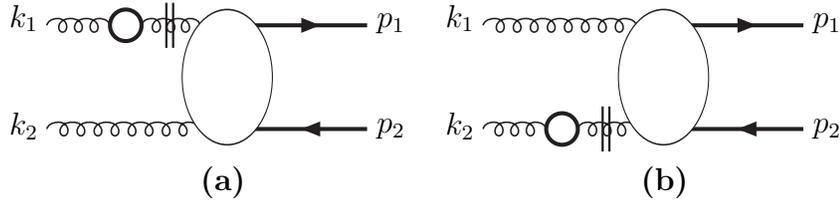

The partonic PDFs and FFs introduced in Sec. \ref{sec:transition} are
given in the $\msbar$ factorization and renormalization scheme.
However, in the FFNS calculations of heavy-quark production
\cite{Nason:1987xz,Nason:1989zy,Beenakker:1988bq,Beenakker:1990ma,
  Bojak:2001fx,Bojak-PhD}, a modification of the $\msbar$ scheme has
been adopted, called $\msbarm$ or decoupling scheme
\cite{Collins:1978wz}, where divergences due to light quarks and gluons
are treated in the $\msbar$ scheme and divergences arising from
heavy-quark loops are subtracted at zero momentum.  In order to switch
from the $\msbarm$ to the $\msbar$ scheme the following terms have to be
added to the partonic cross sections of the fixed-order calculations
(see Sec.\ 3 in Ref.\ \cite{Cacciari:1998it}):
\begin{eqnarray}
-\alpha_s(\mur) \frac{2 T_f}{3 \pi} \ln \frac{\murs}{m^2} 
\der \sigma_{q\qbar}^{(0)}
\qquad (q+\qbar \to Q + X)\, ,
\label{eq:changeqq}
\\
-\alpha_s(\mur) \frac{2 T_f}{3 \pi} \ln \frac{\murs}{\muis} 
\der \sigma_{gg}^{(0)}
\qquad (g+g \to Q+X)\, .
\label{eq:changegg}
\end{eqnarray}
In Eqs.\ \eqref{eq:changeqq} and \eqref{eq:changegg}, the parts
proportional to $\ln \murs$ are due to the change of $\alpha_s$ when
going from the $\msbarm$ to the $\msbar$ scheme.  If
$\alpha_s^{(n_f-1)}(\mur)$ and $\alpha_s^{(n_f)}(\mur)$ denote the
strong-coupling constants in the $\msbarm$ and $\msbar$ schemes,
respectively, one can derive from the renormalization group equation the
following relationship between the couplings:
\begin{equation}
\alpha_s^{(n_f-1)}(\mur)=\alpha_s^{(n_f)}(\mur) 
\left(1- \frac{\alpha_s^{(n_f)}(\mur)}{3\pi}T_f 
\ln \frac{\murs}{m^2}\right) 
+ \Ord(\alpha_s^3)\ .
\end{equation}
The parts proportional to $\ln \muis$ can be obtained by subtracting
from the cross section in the gluon-gluon channel the term
$\left(\fgg(x_1)+\fgg(x_2)\right) \otimes \der \sigmahat^{(0)}(g + g \to
Q + \Qbar)$ (see Fig.\ \ref{fig:cutgg0}).  Since the function $\fgg(x)$
in Eq.\ \eqref{eq:initial} is proportional to Dirac's delta function,
this amounts to a simple multiplication with the Born cross section in
the gluon-gluon channel.  This subtraction term takes into account the
different treatment of heavy-quark loop contributions to external gluon
lines in the $\msbar$ and the $\msbarm$ schemes.  The coefficients in
Ref.\ \cite{Kniehl:2004fy} are given in the $\msbarm$ scheme. Changing
the results in Ref.\ \cite{Kniehl:2004fy} to the $\msbar$ scheme
according to Eqs.\ \eqref{eq:changeqq} and \eqref{eq:changegg} has the
expected effect of replacing $\beta_0^{(n_f-1)}$ by $\beta_0^{(n_f)}$ in
the coefficients $\hat{d}_1$ and $\tilde{d}_1$ in Eqs.\ (28), (29) and
(55) of Ref.\ \cite{Kniehl:2004fy}, so that in the $\msbar$ scheme
$\Delta \hat{d}_1=\Delta \tilde{d}_1=0$ in Eqs.\ (35), (36) and (59) of
Ref.\ \cite{Kniehl:2004fy}.

Even fixing the factorization scheme to the $\msbar$ scheme leaves some
freedom in the implementation of a massive VFNS, as has been discussed
for the case of deep inelastic scattering in Ref.\ \cite{Tung:2001mv}.
Consider for example the condition
\begin{equation}
\lim_{m \to 0}\ (\der \tilde\sigma(m) - \der \sigma^{\sub}(m)) 
\overset{!}{=}
\der \hat{\sigma}_{\msbar} \, ,
\label{eq:condition}
\end{equation}
which might be used as an attempt to define the subtraction terms.
$\der \hat{\sigma}_{\msbar}$ is the massless hard-scattering cross
section in the $\msbar$ scheme. However, this requirement fixes the
subtraction term $\der \sigma^{\sub}(\mu/m, m/p_T)$ only up to terms
$m/p_T$ vanishing in the limit $m \to 0$. The precise treatment of such
terms proportional to $m/p_T$ is not prescribed by factorization.  The
prescription in Eq.\ \eqref{eq:sigmasub}, $\der \sigma^{\sub} = \lim_{m
  \to 0} \der \tilde{\sigma}(m) - \der \hat{\sigma}_{\msbar}$, is
minimal in the sense that no finite mass terms are removed from the hard
part in addition to the collinear logarithms $\ln (m^2/s)$.

The same is true from the point of view of mass factorization: The
factorization and renormalization scheme unambiguously determines the
partonic PDFs and FFs.  However, the convolution prescription leaves
some freedom in the choice of the integration variable and, therefore,
is only unique up to terms of the order $m/Q$ (where $Q$ is the hard
scale).  One example is the ACOT-$\chi$ prescription \cite{Tung:2001mv}
in inclusive DIS, which guarantees the correct threshold behaviour of
the heavy-quark-initiated contributions.  Furthermore, it is possible to
retain the mass terms in the subprocess cross sections entering the
convolution formulas. Actually, this is done in the original ACOT scheme
\cite{Aivazis:1994kh,Aivazis:1994pi}.


\section{Subtraction terms: results} 
\label{sec:subtraction}

We now present the results for the collinear subtraction terms,
calculated using Eqs.\ \eqref{eq:factx1}, \eqref{eq:factx2} and
\eqref{eq:factz}.  The universal partonic PDFs can be found in Sec.\ 
\ref{sec:transition}. The required subprocess cross sections have been
listed for completeness in App.\ \ref{app:subprocesses}.  We retain the
heavy-quark mass terms in the subprocess cross sections entering the
convolutions.  In order to compare with our results in Ref.\ 
\cite{Kniehl:2004fy}, these mass terms have to be dropped.

In order to facilitate the comparison with our previous results, we
expand the subtraction cross section in the following form,
\begin{eqnarray}
\frac{\der^2\sigma^{\sub}}{\der v \der w} 
&=& \Delta c_1 \delta(1-w)+\Delta c_2 
\left (\frac{1}{1-w}\right )_{+}
+ \Delta c_3 \left(\frac{\ln (1-w)}{1-w}\right)_{+} 
\nonumber\\
& &
+ \Delta c_5 \ln v + \Delta c_{10} \ln(1-w) + \Delta c_{11}\, ,
\label{eq:subt}
\end{eqnarray}
and use the abbreviations
\begin{equation}
X = 1-v w, \quad Y=1-v+v w, \quad v_i = i-v \quad (i=1,2)\, .
\end{equation}


\subsection{Subtraction terms for $\ggQ$}
\label{sec:cutgg}

The coefficients $\Delta c_i$ are decomposed into an Abelian and two
non-Abelian parts, following Ref.\ \cite{Bojak-PhD}:
\begin{equation}
\Delta c_i = C(s) \left(C_F^2 \dcqed{i} +\frac{C_A^2}{4} \dcoqu{i} 
+ \frac{1}{4} \dckqu{i}\right) \, ,
\label{eq:colgg}
\end{equation}
with
\begin{equation}
 C(s) = \frac{\alpha_s^3}{2(N^2-1)s}\, .
\end{equation}
There are four different cut diagrams contributing to the total result, 
which we discuss in the following.
\\

\noindent
{\boldmath $\der \sigmahat^{(0)}(g g \to Q \Qbar) \otimes \dcc(z)$}
\\[3mm]
\begin{figure}[t]
\begin{center}
\setlength{\unitlength}{1pt}
\SetScale{0.9}
\begin{picture}(342,63)
\put(0,0){
  \begin{picture}(102,63)
    \SetWidth{0.7}
    \Gluon(0,56)(51,56){4}{5}
    \Gluon(0,7)(51,7){4}{5}
    \Gluon(70,56)(102,31.5){-3}{4}
    \Vertex(51,56){2.0}
    \Vertex(51,7){2.0}
    \Vertex(70,56){2.0}
    \SetWidth{1.8}
    \ArrowLine(102,7)(51,7)
    \ArrowLine(51,7)(51,56)
    \Line(51,56)(76.5,56)
    \ArrowLine(76.5,56)(102,56)
    \SetWidth{0.7}
    \Line(59,62)(59,50)
    \Line(62,62)(62,50)
  \end{picture}
}
\put(120,0){
  \begin{picture}(102,63)
    \SetWidth{0.7}
    \Gluon(0,56)(51,56){4}{5}
    \Gluon(0,7)(51,7){4}{5}
    \Gluon(70,7)(102,31.5){3}{4}
    \Vertex(51,56){2.0}
    \Vertex(51,7){2.0}
    \Vertex(70,7){2.0}
    \SetWidth{1.8}
    \ArrowLine(76.5,7)(102,7)
    \Line(76.5,7)(51,7)
    \ArrowLine(51,56)(51,7)
    \ArrowLine(102,56)(51,56)
    \SetWidth{0.7}
    \Line(59,13)(59,1)
    \Line(62,13)(62,1)
  \end{picture}
}
\put(240,0){
  \begin{picture}(102,63)
    \SetWidth{0.7}
    \Gluon(0,56)(22,31.5){3}{4}
    \Gluon(0,7)(22,31.5){3}{4}
    \Gluon(22,31.5)(67,31.5){3}{5}
    \Gluon(102,30)(88,46.2){3}{3}
    \Vertex(22,31.5){2.0}
    \Vertex(67,31.5){2.0}
    \Vertex(88,46.2){2.0}
    \SetWidth{1.8}
    \ArrowLine(102,7)(67,31.5)
    \ArrowLine(88,46.2)(102,56)
    \Line(67,31.5)(88,46.2)
    \SetWidth{0.7}
    \Line(73,44)(82,34)
    \Line(70,42)(79,32)
  \end{picture}
}
\end{picture}
\end{center}
\caption{Feynman diagrams representing $\der \sigmahat^{(0)}(g g \to Q
  \Qbar) \otimes \dcc(z)$.} 
\label{fig:cutgg1}
\end{figure}
\noindent
The cut diagrams are shown in Fig.\ \protect\ref{fig:cutgg1}.  The cross
section $\der \sigmahat^{(0)}(g g \to Q \Qbar)$ is proportional to the
function
\begin{equation}
\tau(v) = \frac{v}{1-v} + \frac{1-v}{v} 
+ \frac{4 m^2}{s v (1-v)} 
\left(1-\frac{m^2}{s v (1-v)}\right)\, ,
\end{equation}
which appears in the following expressions for the $\Delta c_i$.  They
are given by
\begin{eqnarray}
\Delta c_1 & = & \left[\left(\frac{3}{4} + \ln v\right)\Lfin 
+ 1-\ln v -\ln ^2 v\right]
\times
2 C(s) C_F \tau (v) [C_F -C_A v(1-v)]\, ,
\label{eq:dc1gg}
\\[1.5ex]
\Delta c_2 & = &
\left(\Lfin - 1 - 2 \ln v \right) 
\times
2 C(s) C_F \tau(v) [C_F -C_A v(1-v)]\, ,
\label{eq:dc2gg}
\\[1.5ex]
\Delta c_3 & = & -2 
\times 
2 C(s) C_F \tau(v) \left[C_F -C_A v(1-v) \right]\, ,
\label{eq:dc3gg}
\\[1.5ex]
\Delta c_5 & = & C(s) \left(C_F^2 \dcqed{5} + \frac{1}{4}(C_A^2-1) 
\dcoqu{5} \right) \, ,
\label{eq:dc5gg}
\end{eqnarray}
where
\begin{eqnarray}
\dcqed{5} &=&
\frac{2 v}{v_1} - 
\frac{2 ( 2 - 2 v + v^2 ) }{v w} + 
\frac{2 v^2 w}{v_1} + \frac{4 v}{Y}
\nonumber\\
& &
+ \frac{m^2}{s} \left(
\frac{8 v ( 3 - 2 v ) }{v_1} - 
\frac{8 ( 2 - 2 v + v^2 ) }{ v w} + 
\frac{8 v^2 w}{v_1}
\right)
\nonumber\\
& &
+\frac{m^4}{s^2} \bigg(
\frac{-8 v ( 11 - 15 v + 6 v^2 ) }{v_1^2} +
\frac{8 v_1 ( 2 - 2 v + v^2 ) }{v^2 w^2} 
\nonumber\\
& &
\phantom{+\frac{m^4}{s^2}\bigg(}
+ \frac{8 ( 2 + 4 v - 7 v^2 + 4 v^3 ) }{v^2 w} 
+ \frac{8 v^2 ( -5 + 4 v )  w}{v_1^2} 
- \frac{8 v^3 w^2}{v_1^2}
\bigg)\, ,
\\
\nonumber\\
\dcoqu{5} &=&
-4 v 
+ \frac{8 v v_1^2 }{Y^3} 
- \frac{8 v^2 v_1}{Y^2} 
+ \frac{4 v ( 3  - 6 v + 4 v^2 ) }{Y}
+
\frac{m^2}{s} \left(-16 v + \frac{16 v}{Y}\right)
\nonumber\\
& &
+
\frac{m^4}{s^2} \left(
\frac{16 v ( 3 - 2 v ) }{v_1} - 
  \frac{16 ( 2 - 2 v + v^2 ) }{v w} +
  \frac{16 v^2 w}{v_1}
\right)\, .
\end{eqnarray}
We find that $\dckqu{5} = - \dcoqu{5}$ and, finally, 
\begin{eqnarray}
\Delta c_{10} &=& \Delta c_5 \, ,
\label{eq:dc10gg}
\\
\Delta c_{11} &=& \tfrac{1}{2} \Delta c_5 \left(1 - \Lfin\right)\ .
\label{eq:dc11gg}
\end{eqnarray}
The latter two relations can be derived from $\dcc(\zb$) in Eq.\ 
\eqref{eq:dcc} inspecting the expressions for $B(v)$, $C(v)$ and
$D(v,w)$.  (Note that the parts with $B(v)$ and $C(v)$ contribute to
$\Delta c_{11}$ and $\Delta c_{10}$, respectively, in cases where the
$1/(1-w)_{+}$ is canceled by factors $(1-w)$.)

Now we turn to a comparison with the results in Ref.\ 
\cite{Kniehl:2004fy}, which have been derived as described in Sec.\ 
\ref{sec:hadro}.  For $\muf = m$ (and neglecting terms of the order
$\Ord(m^2/s)$), Eqs.\ \eqref{eq:dc1gg}--\eqref{eq:dc11gg} are in
complete agreement with Eqs.\ (18) and (21)--(24) in Ref.\ 
\cite{Kniehl:2004fy}.  Furthermore, the parts proportional to $\ln
(\mufs/m^2)$ in Eqs.\ \eqref{eq:dc1gg} and \eqref{eq:dc2gg} are
identical to Eqs.\ (37) and (38) in Ref.\ \cite{Kniehl:2004fy}.  As for
$\Delta c_{11}$ in Eq.\ \eqref{eq:dc11gg}, the part proportional to $\ln
(\mufs/m^2)$ is in agreement with Eqs.\ (41) and (43) for the 'qed' and
'kq' parts. The 'oq' part reproduces Eq.\ (42) in Ref.\ 
\cite{Kniehl:2004fy} after adding the contribution given in Eq.\ 
\eqref{eq:dc11oq}.
\\


\noindent
{\boldmath $\der \sigmahat^{(0)}(g g \to g g) \otimes \dcg(z)$}
\\[3mm]
\begin{figure}[t]
\begin{center}
\setlength{\unitlength}{1pt}
\SetScale{0.8}
\begin{picture}(432,63)
\put(0,0){
  \begin{picture}(102,63)
    \SetWidth{0.7}
    \Gluon(0,56)(15.75,43.75){3}{2}
    \Gluon(15.75,43.75)(31.5,31.5){3}{2}
    \Gluon(0,7)(31.5,31.5){3}{4}
    \Gluon(31.5,31.5)(80,31.5){3}{6}
    \Gluon(15.75,43.75)(75.5,56){3}{7}
    \Vertex(15.75,43.75){2.0}
    \Vertex(31.5,31.5){2.0}
    \Vertex(80,31.5){2.0}
    \SetWidth{1.8}
    \ArrowLine(102,7)(80,31.5)
    \ArrowLine(80,31.5)(102,56)
    \SetWidth{1.0}
    \Line(69,43)(69,20)
    \Line(73,43)(73,20)
  \end{picture}
}
\put(110,0){
  \begin{picture}(102,63)
    \SetWidth{0.7}
    \Gluon(0,56)(31.5,31.5){3}{4}
    \Gluon(15.75,19.25)(31.5,31.5){3}{2}
    \Gluon(0,7)(15.75,19.25){3}{2}
    \Gluon(31.5,31.5)(80,31.5){3}{6}
    \Gluon(75.5,7)(15.75,19.25){3}{7}
    \Vertex(15.75,19.25){2.0}
    \Vertex(31.5,31.5){2.0}
    \Vertex(80,31.5){2.0}
    \SetWidth{1.8}
    \ArrowLine(102,7)(80,31.5)
    \ArrowLine(80,31.5)(102,56)
    \SetWidth{1.0}
    \Line(69,43)(69,20)
    \Line(73,43)(73,20)
  \end{picture}
}
\put(220,0){
  \begin{picture}(102,63)
    \SetWidth{0.7}
    \Gluon(0,56)(22,31.5){3}{4}
    \Gluon(0,7)(22,31.5){3}{4}
    \Gluon(22,31.5)(80,31.5){3}{6}
    \Gluon(22,31.5)(46.5,56){3}{3}
    \Vertex(22,31.5){2.0}
    \Vertex(80,31.5){2.0}
    \SetWidth{1.8}
    \ArrowLine(102,7)(80,31.5)
    \ArrowLine(80,31.5)(102,56)
    \SetWidth{1.0}
    \Line(69,43)(69,20)
    \Line(73,43)(73,20)
  \end{picture}
}
\put(330,0){
  \begin{picture}(102,63)
    \SetWidth{0.7}
    \Gluon(0,56)(22,31.5){3}{4}
    \Gluon(0,7)(22,31.5){3}{4}
    \Gluon(22,31.5)(51,31.5){3}{3}
    \Gluon(51,31.5)(80,31.5){3}{3}
    \Line(51,31.5)(51,33.5)
    \Gluon(51,33.5)(75.5,56){3}{3}
    \Vertex(22,31.5){2.0}
    \Vertex(51,31.5){2.0}
    \Vertex(80,31.5){2.0}
    \SetWidth{1.8}
    \ArrowLine(102,7)(80,31.5)
    \ArrowLine(80,31.5)(102,56)
    \SetWidth{1.0}
    \Line(69,43)(69,20)
    \Line(73,43)(73,20)
  \end{picture}
}
\end{picture}
\end{center}
\caption{Feynman diagrams representing of $\der \sigmahat^{(0)}(g g \to
  g g) \otimes \dcg(z)$.} 
\label{fig:cutgg2}
\end{figure}
\noindent
The cut diagram Fig.\ \ref{fig:cutgg2} only contributes to the part of
$\Delta c_{11}$ proportional to $C_A^2$:
\begin{eqnarray}
\dcoqu{11} &=& \bigg(
-48 v^2 
+ \frac{8 v_1  ( 1 - 2 v + 2 v^2 ) }{v w^2} 
+ \frac{16 ( 1 - 3 v + 2 v^2 ) }{w} 
\nonumber\\
& &
+ \frac{8 v^2 ( 7 - 14 v + 8 v^2 )  w}{v_1^2} 
- \frac{16 v^3 ( -1 + 2 v )  w^2}{v_1^2} 
+ \frac{16 v^4 w^3}{v_1^2} 
\nonumber\\
& &
+ \frac{8 v v_1^2 }{Y^3} 
- \frac{8 ( 3 v - 5 v^2 + 2 v^3 ) }{Y^2} 
+ \frac{8 ( 7 v - 6 v^2 + 2 v^3 ) }{Y}
\bigg) \Lfin\, .
\label{eq:dc11oq}
\end{eqnarray}


\noindent
{\boldmath $\fcg(x_1) \otimes \der \sigmahat^{(0)}(Q g \to Q g)$}
\\[3mm]
\noindent
The $u$-channel cut in the initial state described by the diagram in
Fig.\ \ref{fig:cutgg3} contributes:
\begin{figure}[t]
\begin{center}
\setlength{\unitlength}{1pt}
\SetScale{0.9}
\begin{picture}(342,63)
\put(0,0){
  \begin{picture}(102,63)
    \SetWidth{0.7}
    \Gluon(0,56)(51,56){4}{5}
    \Gluon(0,7)(51,7){4}{5}
    \Gluon(51,31.5)(102,31.5){3}{6}
    \Vertex(51,56){2.0}
    \Vertex(51,7){2.0}
    \Vertex(51,31.5){2.0}
    \SetWidth{1.8}
    \ArrowLine(51,7)(102,7)
    \ArrowLine(51,31.5)(51,7)
    \Line(51,56)(51,31.5)
    \ArrowLine(102,56)(51,56)
    \SetWidth{1.0}
    \Line(43,46)(59,46)
    \Line(43,43)(59,43)
  \end{picture}
}
\put(120,0){
  \begin{picture}(102,63)
    \SetWidth{0.7}
    \Gluon(0,56)(51,56){4}{5}
    \Gluon(0,7)(51,7){4}{5}
    \Gluon(70,7)(102,31.5){3}{4}
    \Vertex(51,56){2.0}
    \Vertex(51,7){2.0}
    \Vertex(70,7){2.0}
    \SetWidth{1.8}
    \ArrowLine(76.5,7)(102,7)
    \Line(51,7)(76.5,7)
    \ArrowLine(51,56)(51,7)
    \ArrowLine(102,56)(51,56)
    \SetWidth{1.0}
    \Line(43,46)(59,46)
    \Line(43,43)(59,43)
  \end{picture}
}
\put(240,0){
  \begin{picture}(102,63)
    \SetWidth{0.7}
    \Gluon(0,56)(51,56){4}{5}
    \Gluon(0,7)(51,7){3}{7}
    \Gluon(51,7)(102,7){3}{7}
    \Line(51,12)(51,7)
    \Gluon(51,12)(51,31.5){3}{3}
    \Vertex(51,56){2.0}
    \Vertex(51,7){2.0}
    \Vertex(51,31.5){2.0}
    \SetWidth{1.8}
    \ArrowLine(51,31.5)(102,31.5)
    \Line(51,56)(51,31.5)
    \ArrowLine(102,56)(51,56)
    \SetWidth{1.0}
    \Line(43,46)(59,46)
    \Line(43,43)(59,43)
  \end{picture}
}
\end{picture}
\end{center}
\caption{Feynman diagrams representing of $\fcg(x_1) \otimes \der
  \sigmahat^{(0)}(Q g \to Q g)$.} 
\label{fig:cutgg3}
\end{figure}
\begin{eqnarray}
\dcqed{11} &=& 
\label{eq:dc11qed3}
\frac{1+v^2}{v w} (1-2 w + 2 w^2) \Lini \, ,
\\
\dcoqu{11} &=& 
\frac{1+v^2}{v_1^2}\frac{2}{w} (1-2 w + 2 w^2) \Lini  \, ,
\label{eq:dc11oqu3}
\\
\dckqu{11} &=& - \dcoqu{11}\, .
\label{eq:dc11kqu3}
\end{eqnarray}


\noindent
{\boldmath $\fcg(x_2) \otimes \der \sigmahat^{(0)}(g Q \to Q g)$}
\\[3mm]
\noindent
Finally, we have a contribution from the $u$-channel cut in the initial
state described by the diagram in Fig.\ \ref{fig:cutgg4}:
\begin{figure}[t]
\begin{center}
\setlength{\unitlength}{1pt}
\SetScale{0.9}
\begin{picture}(342,63)
\put(0,0){
  \begin{picture}(102,63)
    \SetWidth{0.7}
    \Gluon(0,56)(51,56){4}{5}
    \Gluon(0,7)(51,7){4}{5}
    \Gluon(51,31.5)(102,31.5){3}{6}
    \Vertex(51,56){2.0}
    \Vertex(51,7){2.0}
    \Vertex(51,31.5){2.0}
    \SetWidth{1.8}
    \ArrowLine(102,7)(51,7)
    \Line(51,7)(51,31.5)
    \ArrowLine(51,31.5)(51,56)
    \ArrowLine(51,56)(102,56)
    \SetWidth{1.0}
    \Line(43,22)(59,22)
    \Line(43,19)(59,19)
  \end{picture}
}
\put(120,0){
  \begin{picture}(102,63)
    \SetWidth{0.7}
    \Gluon(0,56)(51,56){4}{5}
    \Gluon(0,7)(51,7){4}{5}
    \Gluon(70,56)(102,31.5){-3}{4}
    \Vertex(51,56){2.0}
    \Vertex(51,7){2.0}
    \Vertex(70,56){2.0}
    \SetWidth{1.8}
    \ArrowLine(102,7)(51,7)
    \ArrowLine(51,7)(51,56)
    \Line(51,56)(76.5,56)
    \ArrowLine(76.5,56)(102,56)
    \SetWidth{1.0}
    \Line(43,22)(59,22)
    \Line(43,19)(59,19)
  \end{picture}
}
\put(240,0){
  \begin{picture}(102,63)
    \SetWidth{0.7}
    \Gluon(0,7)(51,7){4}{5}
    \Gluon(0,56)(51,56){3}{7}
    \Gluon(51,56)(102,56){3}{7}
    \Line(51,56)(51,51)
    \Gluon(51,31.5)(51,51){3}{3}
    \Vertex(51,56){2.0}
    \Vertex(51,7){2.0}
    \Vertex(51,31.5){2.0}
    \SetWidth{1.8}
    \ArrowLine(51,31.5)(102,31.5)
    \Line(51,7)(51,31.5)
    \ArrowLine(102,7)(51,7)
    \SetWidth{1.0}
    \Line(43,22)(59,22)
    \Line(43,19)(59,19)
  \end{picture}
}
\end{picture}
\end{center}
\caption{Feynman diagrams representing of $\fcg(x_2) \otimes \der
  \sigmahat^{(0)}(g Q \to Q g)$.} 
\label{fig:cutgg4}
\end{figure}
\begin{eqnarray}
\label{eq:dc11qed4}
\dcqed{11} &=& 
\bigg(
\frac{v ( -1 + 2 v ) }{v_1}   
- \frac{v^2 w}{v_1} 
+ \frac{2 v_1  v}{X^3} 
- \frac{2 v}{X^2} 
+ \frac{v ( 3 - 4 v + 2 v^2 ) }{v_1 X}
\bigg) \Lini  \, ,
\\
\dcoqu{11} &=& \bigg(
\frac{2 v}{v_1} 
+ \frac{4 ( 1 - 2 v + 2 v^2 ) }{v_1  v w^2} 
+ \frac{4 ( 1 - 4 v + 2 v^2 ) }{v_1  w} 
+ \frac{4 v v_1 }{X^2} 
+ \frac{4 v( 1 - 2 v ) }{X}
\bigg) \times
\label{eq:dc11oqu4}
\nonumber\\
& &
\Lini\, ,
\\
\dckqu{11} &=& - \dcoqu{11}\, .
\label{eq:dc11kqu4}
\end{eqnarray}
The sum of Eqs.\ \eqref{eq:dc11qed3} and \eqref{eq:dc11qed4}, that of
Eqs.\ \eqref{eq:dc11oqu3} and \eqref{eq:dc11oqu4}, and that of Eqs.\ 
\eqref{eq:dc11kqu3} and \eqref{eq:dc11kqu4} are identical to Eqs.\ (44),
(45), and (46) in Ref.\ \cite{Kniehl:2004fy}, respectively.


\subsection{Subtraction terms for $\qqbQ$}
\label{sec:cutqq}

The results for the coefficients $\Delta c_i$ have the following colour
decomposition: 
\begin{equation}
\Delta c_i = C_q(s) \frac{C_F}{2} 
\left(C_F \dccf{i} +C_A \dcca{i}\right) \, ,
\label{eq:colqq}
\end{equation} 
with
\begin{equation}
C_q(s) = \frac{\alpha_s^3}{2 N s} \, .
\end{equation}
There are two different cut diagrams contributing to the total result.
\\


\noindent
{\boldmath $\der \sigmahat^{(0)}(q \qbar \to Q \Qbar) \otimes \dcc(z)$}
\\[3mm]
\begin{figure}[t]
\begin{center}
\setlength{\unitlength}{1pt}
\SetScale{0.9}
  \begin{picture}(102,63)
    \SetWidth{0.7}
    \ArrowLine(0,56)(22,31.5)
    \ArrowLine(22,31.5)(0,7)
    \Gluon(22,31.5)(67,31.5){3}{5}
    \Gluon(102,30)(88,46.2){3}{3}
    \Vertex(22,31.5){2.0}
    \Vertex(67,31.5){2.0}
    \Vertex(88,46.2){2.0}
    \SetWidth{1.8}
    \ArrowLine(102,7)(67,31.5)
    \ArrowLine(88,46.2)(102,56)
    \Line(67,31.5)(88,46.2)
    \SetWidth{0.7}
    \Line(73,44)(82,34)
    \Line(70,42)(79,32)
  \end{picture}
\end{center}
\caption{Feynman diagrams representing of $\der \sigmahat^{(0)}(q \qbar
  \to Q \Qbar)\otimes \dcc(z)$.} 
\label{fig:cutqq1}
\end{figure}
\noindent
Figure \ref{fig:cutqq1} shows the diagram with a cut in the final state.
The $2 \to 2$ process cross section $\der \sigmahat(q \qbar \to Q
\Qbar)$ is proportional to the function
\begin{equation}
\tauq = (1-v)^2+v^2 + \frac{2 m^2}{s} \, ,
\end{equation}
which will occur in the results below.  Furthermore, for this
contribution, the $C_A$ parts vanish, i.e., $\Delta c_i^{\rm ca}=0$
($i=1,2,3,5,10,11$).  Therefore, the final results are proportional to
the colour factor $C_F^2$:
\begin{eqnarray}
\label{eq:dc1qq}
\Delta c_1 &=&
\left[
\left(\frac{3}{4} + \ln v\right)\Lfin + 1-\ln v - \ln^2 v
\right]
\times 2 C_q(s)\tauq C_F^2\, ,
\\
\nonumber\\
\label{eq:dc2qq}
\Delta c_2 &=&
\left( \Lfin - 1 - 2 \ln v \right) 
\times 2 C_q(s)\tauq C_F^2\, ,
\\
\nonumber\\
\label{eq:dc3qq}
\Delta c_3 &=& -2 \times 2 C_q(s)\tauq C_F^2\, ,
\\
\nonumber\\
\Delta c_5 &=& 
2 C_q(s) C_F^2 \left[
v - \frac{2 v v_1^2 }{Y^3} 
+ \frac{2 v^2 v_1}{Y^2} 
- \frac{3 v - 6 v^2 + 4 v^3}{Y}
+ \frac{m^2}{s} 2 v \left(1 - \frac{1}{Y}\right)
\right]\, . 
\label{eq:dc5qq}
\end{eqnarray}
Finally, we find again 
\begin{eqnarray}
\label{eq:dc10qq}
\Delta c_{10} &=& \Delta c_5\, ,
\\
\Delta c_{11} &=& \tfrac{1}{2} \Delta c_5 \left(1 - \Lfin \right)\ .
\label{eq:dc11qq}
\end{eqnarray}
One can observe the same structure of the results as for $\der
\sigmahat^{(0)}(g g \to Q \Qbar) \otimes \dcc(z)$ given in Sec.\ 
\ref{sec:cutgg}.

Now we turn again to the comparison with the results in Ref.\ 
\cite{Kniehl:2004fy}.  For $\muf = m$ (and neglecting terms of the order
$\Ord(m^2/s)$), Eqs.\ \eqref{eq:dc1qq}--\eqref{eq:dc11qq} are identical
to Eqs.\ (51)--(54) in Ref.\ \cite{Kniehl:2004fy}.  The parts
proportional to $\ln (\mufs/m^2)$ in Eqs.\ \eqref{eq:dc1qq} and
\eqref{eq:dc2qq} coincide with Eqs.\ (60) and (61) in Ref.\ 
\cite{Kniehl:2004fy}.  As for $\Delta c_{11}$ in Eq.\ \eqref{eq:dc11qq},
the part proportional to $\ln (\mufs/m^2)$ is in agreement with Eq.\ 
(62) in Ref.\ \cite{Kniehl:2004fy} only after including the contribution
from $\der \sigmahat(q \qbar \to g g) \otimes \dcg(z)$, which will be
given in the next subsection.
\\


\noindent
{\boldmath $\der \sigmahat^{(0)}(q \qbar \to g g) \otimes \dcg(z)$}
\\[3mm]
\begin{figure}[t]
\begin{center}
\setlength{\unitlength}{1pt}
\SetScale{0.9}
\begin{picture}(342,63)
\put(0,0){
  \begin{picture}(102,63)
    \SetWidth{0.7}
    \ArrowLine(0,56)(15.75,43.75)
    \ArrowLine(15.75,43.75)(31.5,31.5)
    \ArrowLine(31.5,31.5)(0,7)
    \Gluon(31.5,31.5)(80,31.5){3}{6}
    \Gluon(15.75,43.75)(75.5,56){3}{7}
    \Vertex(15.75,43.75){2.0}
    \Vertex(31.5,31.5){2.0}
    \Vertex(80,31.5){2.0}
    \SetWidth{1.8}
    \ArrowLine(102,7)(80,31.5)
    \ArrowLine(80,31.5)(102,56)
    \SetWidth{1.0}
    \Line(69,43)(69,20)
    \Line(73,43)(73,20)
  \end{picture}
}
\put(120,0){
  \begin{picture}(102,63)
    \SetWidth{0.7}
    \ArrowLine(0,56)(31.5,31.5)
    \ArrowLine(31.5,31.5)(15.75,19.25)
    \ArrowLine(15.75,19.25)(0,7)
    \Gluon(31.5,31.5)(80,31.5){3}{6}
    \Gluon(75.5,7)(15.75,19.25){3}{7}
    \Vertex(15.75,19.25){2.0}
    \Vertex(31.5,31.5){2.0}
    \Vertex(80,31.5){2.0}
    \SetWidth{1.8}
    \ArrowLine(102,7)(80,31.5)
    \ArrowLine(80,31.5)(102,56)
    \SetWidth{1.0}
    \Line(69,43)(69,20)
    \Line(73,43)(73,20)
  \end{picture}
}
\put(240,0){
  \begin{picture}(102,63)
    \SetWidth{0.7}
    \ArrowLine(0,56)(22,31.5)
    \ArrowLine(22,31.5)(0,7)
    \Gluon(22,31.5)(51,31.5){3}{3}
    \Gluon(51,31.5)(80,31.5){3}{3}
    \Line(51,31.5)(51,33.5)
    \Gluon(51,33.5)(75.5,56){3}{3}
    \Vertex(51,31.5){2.0}
    \Vertex(22,31.5){2.0}
    \Vertex(80,31.5){2.0}
    \SetWidth{1.8}
    \ArrowLine(102,7)(80,31.5)
    \ArrowLine(80,31.5)(102,56)
    \SetWidth{1.0}
    \Line(69,43)(69,20)
    \Line(73,43)(73,20)
  \end{picture}
}
\end{picture}
\end{center}
\caption{Feynman diagrams representing of $\der \sigmahat^{(0)}(q \qbar
  \to g g) \otimes \dcg(z)$.} 
\label{fig:cutqq2}
\end{figure}
\noindent
The result for the cut diagram Fig.\ \ref{fig:cutqq2} reads
\begin{eqnarray}
\label{eq:dccf11qq}
\dccf{11} & = &
\left[
\frac{2 v ( 3 - 4 v + 2 v^2 ) }{v_1} 
+ \frac{2(1 - 2 v + 2 v^2)}{w} 
- \frac{4 v^3 w}{v_1} 
+ \frac{4 v^3 w^2}{v_1} 
- \frac{4 v}{Y}
\right]
\Lfin\, ,
\\
\dcca{11} & = &
\left[
4 (2-v)  v - 4 v^2 w 
- \frac{4 v v_1^2 }{Y^3} 
+ \frac{4 ( 3 v - 5 v^2 + 2 v^3 ) }{Y^2} 
- \frac{2(9 v - 12 v^2 + 4 v^3)}{Y}
\right] \times
\nonumber\\
& &
\Lfin\, .
\label{eq:dcca11qq}
\end{eqnarray}
Since $\Delta c_{11}$ in Eq.\ \eqref{eq:dc11qq} only has a $C_F$ part,
Eq.\ \eqref{eq:dcca11qq} is the only contribution to the $C_A$ part and
hence is in agreement with Eq.\ (64) in Ref.\ \cite{Kniehl:2004fy}.
Furthermore, it is easy to see that the sum of $\dccf{11}$ in Eq.\ 
\eqref{eq:dc11qq}, taken from the part $\propto \ln (\mufs/m^2)$, and
$\dccf{11}$ in Eq.\ \eqref{eq:dccf11qq} reproduces Eq.\ (63) in Ref.\ 
\cite{Kniehl:2004fy}.


\subsection{Subtraction terms for $\gqQ$}
\label{sec:cutgq}

The process $g q \to Q \Qbar q$ appears for the first time at NLO. It
has the colour decomposition
\begin{equation}
\Delta c_i = \Cgq \left(C_F \dccf{i} + C_A \dcca{i} \right) \, ,
\label{eq:colgq}
\end{equation} 
with
\begin{equation}
\Cgq = \frac{\alpha_s^3}{2 N s}\, .
\end{equation}
There are two different cut diagrams contributing to the total result.
The results for the process $g \qbar \to Q \Qbar \qbar$ are the same as
can be easily seen with the help of the expressions in App.\ 
\ref{app:cutgq}.
\\


\noindent
{\boldmath $\der \sigmahat^{(0)}(g q \to g q) \otimes \dcg(z)$}
\\[3mm]
\begin{figure}[t!]
\begin{center}
\setlength{\unitlength}{1pt}
\SetScale{0.9}
\begin{picture}(342,65)
\put(0,0){
  \begin{picture}(102,63)
    \SetWidth{0.7}
    \Gluon(0,56)(32,20){3}{5}
    \Gluon(70,20)(80,43.75){3}{3}
    \ArrowLine(0,7)(32,20)
    \ArrowLine(32,20)(70,20)
    \ArrowLine(70,20)(102,7)
    \Vertex(32,20){2.0}
    \Vertex(70,20){2.0}
    \Vertex(80,43.75){2.0}
    \SetWidth{1.8}
    \ArrowLine(102,31.5)(80,43.75)
    \ArrowLine(80,43.75)(102,56)
    \SetWidth{1.0}
    \Line(66,39)(84,30)
    \Line(65,35)(83,26)
  \end{picture}
}
\put(120,0){
  \begin{picture}(102,63)
    \SetWidth{0.7}
    \Gluon(0,56)(44.7,33){3}{5}
    \Gluon(54.4,28)(70,20){-3}{1}
    \Gluon(32,20)(80,43.75){3}{5}
    \ArrowLine(0,7)(32,20)
    \ArrowLine(32,20)(70,20)
    \ArrowLine(70,20)(102,7)
    \Vertex(32,20){2.0}
    \Vertex(70,20){2.0}
    \Vertex(80,43.75){2.0}
    \SetWidth{1.8}
    \ArrowLine(102,31.5)(80,43.75)
    \ArrowLine(80,43.75)(102,56)
    \SetWidth{1.0}
    \Line(61,48)(80,34)
    \Line(57,46)(76,32)
  \end{picture}
}
\put(240,0){
\begin{picture}(102,65)
  \SetWidth{0.7}
  \Gluon(0,56)(51,31.5){3}{5}
  \ArrowLine(0,7)(51,7)
  \ArrowLine(51,7)(102,7)
  \Gluon(51,7)(51,31.5){3}{3}
  \Gluon(51,31.5)(80,43.75){3}{3}
  \Vertex(51,7){2.0}
  \Vertex(51,31.5){2.0}
  \Vertex(80,43.75){2.0}
  \SetWidth{1.8}
  \ArrowLine(102,31.5)(80,43.75)
  \ArrowLine(80,43.75)(102,56)
    \SetWidth{1.0}
    \Line(61,48)(80,34)
    \Line(57,46)(76,32)
\end{picture}
}
\end{picture}
\end{center}
\caption{Feynman diagrams representing of $\der \sigmahat^{(0)}(g q \to
  g q) \otimes \dcg(z)$.} 
\label{fig:cutgq1}
\end{figure}
\noindent
For the cut diagram shown Fig.\ \ref{fig:cutgq1}, we find
\begin{eqnarray}
\label{eq:dccf11gq}
\dccf{11} &=&
\left(
-2 v^2 + \frac{1 - 2 v + 2 v^2}{2 w} + 2 v^2 w 
- \frac{v_1  v}{2 Y^2} 
+ \frac{3 v - 2 v^2}{2 Y}
\right) \Lfin\, ,
\\
\dcca{11} &=&
\left(
-v^2 
+ \frac{v^2 ( 2 - 4 v + 3 v^2 )  w}{v_1^2} 
+ \frac{2 v^3 ( 1 - 2 v )  w^2}{v_1^2} 
+ \frac{2 v^4 w^3}{v_1^2} 
+ \frac{v}{2 Y}
\right) \Lfin\, .
\label{eq:dcca11gq}
\end{eqnarray}
Equations \eqref{eq:dccf11gq} and \eqref{eq:dcca11gq} are in agreement
with Eqs.\ (69) and (70) in Ref.\ \cite{Kniehl:2004fy}.
\\[3mm]


\noindent
{\boldmath $\fcg(x_1) \otimes \der \sigmahat^{(0)}(Q q \to Q q)$}
\\[3mm]
\begin{figure}[t]
\begin{center}
\setlength{\unitlength}{1pt}
\SetScale{0.9}
  \begin{picture}(102,63)
    \SetWidth{0.7}
    \Gluon(0,56)(51,56){4}{5}
    \ArrowLine(0,7)(51,7)
    \ArrowLine(51,7)(102,7)
    \Gluon(51,7)(51,31.5){3}{3}
    \Vertex(51,56){2.0}
    \Vertex(51,7){2.0}
    \Vertex(51,31.5){2.0}
    \SetWidth{1.8}
    \ArrowLine(51,31.5)(102,31.5)
    \Line(51,56)(51,31.5)
    \ArrowLine(102,56)(51,56)
    \SetWidth{1.0}
    \Line(44,46)(58,46)
    \Line(44,42)(58,42)
  \end{picture}
\end{center}
\caption{Feynman diagrams representing of $\fcg(x_1) \otimes \der
  \sigmahat^{(0)}(Q q \to Q q)$.} 
\label{fig:cutgq2}
\end{figure}
\noindent
The contribution of the cut diagram Fig.\ \ref{fig:cutgq2} is given by
\begin{eqnarray}
\label{eq:dccf11gq2}
\dccf{11} &=&
\frac{1 + v^2}{2 v_1^2 w} \left(1 -2 w+2 w^2\right)\Lini\, ,
\\
\dcca{11} &=& 0\, .
\end{eqnarray}
Equation \eqref{eq:dccf11gq2} is identical to Eq.\ (68) in Ref.\ 
\cite{Kniehl:2004fy}.


\section{Conclusions and discussion}
\label{sec:summary}

\begin{table}[t]
\begin{center}
\begin{tabular}{l|l|l|l}
channel & & this paper & Ref.\ \protect\cite{Kniehl:2004fy}\\ 
\hline
{\begin{minipage}{2.20cm}
$g g \to Q \Qbar g$:\end{minipage}} & &  & \\
& $\muf = m$ & \eqref{eq:dc1gg}--\eqref{eq:dc11gg} & (18), (21)--(24) \\ 
& $\propto \Lfin$ & \eqref{eq:dc1gg}, \eqref{eq:dc2gg} & (37), (38) \\
& $\propto \Lfin$ & \eqref{eq:dc11gg}[QED-part], \eqref{eq:dc11gg}[KQ-part]
& (41), (43)\\
& $\propto \Lfin$ & \eqref{eq:dc11gg}[OQ-part]+\eqref{eq:dc11oq}
& (42)\\
& $\propto \Lini$ & \eqref{eq:dc11qed3}+\eqref{eq:dc11qed4},
\eqref{eq:dc11oqu3}+\eqref{eq:dc11oqu4},
\eqref{eq:dc11kqu3}+\eqref{eq:dc11kqu4}
& (44), (45), (46) \\
& & & \\
$q \qbar \to Q \Qbar g$: & &  & \\
& $\muf = m$ & \eqref{eq:dc1qq}--\eqref{eq:dc11qq} & (51)--(54) \\ 
& $\propto \Lfin$ & \eqref{eq:dc1qq}, \eqref{eq:dc2qq} & (60), (61) \\
& $\propto \Lfin$ & \eqref{eq:dc11qq}[$C_F$-part] + \eqref{eq:dccf11qq} 
& (63)\\
& $\propto \Lfin$ & \eqref{eq:dcca11qq} & (64) \\
& & & \\
$g q \to Q \Qbar q$: & &  & \\
& $\propto \Lfin$ & \eqref{eq:dccf11gq}, \eqref{eq:dcca11gq} & (69), (70) \\
& $\propto \Lini$ & \eqref{eq:dccf11gq2} & (68)
\end{tabular}
\end{center}

\caption{Collinear subtraction terms for the partonic subprocesses
  $\ggQ$, $\qqbQ$ and $\gqQ$ in comparison with the results of Ref.\
  \protect\cite{Kniehl:2004fy}. In the second column, '$\muf = m$'
  indicates that the final state factorization scale $\muf$ has to be
  set to $m$ in the equations in the third column. Furthermore, $\propto
  \Lfin$ ($\propto \Lini$) refers to those parts of the equations in the
  third column which are proportional to $\Lfin$ ($\Lini$). The third
  and forth columns list the equation numbers for the corresponding
  subtraction terms derived in this paper and in Ref.\
  \protect\cite{Kniehl:2004fy}, respectively. Equations combined by a 
  'plus' sign have to be added. 
}
\label{tab:comparison}
\end{table}
We have presented a detailed description of the derivation of collinear
subtraction terms with the help of mass factorization keeping the
heavy-quark mass as a regulator for collinear divergences.  As an
example, we have considered heavy-quark production in hadron-hadron
collisions, which is the most complex case.  With one minor exception
(see below), we have reproduced all the subtraction terms derived in
Ref.\ \cite{Kniehl:2004fy} and found nice agreement.  For a summary of
the comparison, see Table \ref{tab:comparison}.  Apart from giving
additional insight and providing a consistency check of our previous
results, this detailed example will be useful for extending the GM-VFN
scheme to other processes.

Note however, that some exceptions have been found.
(i) In Eq. (25) of Ref.\ \cite{Kniehl:2004fy}, we have found an extra
contribution to the coefficient $\Delta c_1$ in the $gg \to Q\Qbar$
channel resulting in a modification $\Delta c_1 \to \Delta c_1 - C(s)
C_A \tfrac{1}{9} v(1-v)$.  This extra piece has its origin in
heavy-quark loop contributions to the virtual corrections to the Born
process $gg \to Q\Qbar$ in Ref.\ \cite{Bojak:2001fx} and has no
counterpart in the results of Sec.\ \ref{sec:cutgg}.  However, these
terms are absent in the massless limit of the calculation in Refs.\ 
\cite{Beenakker:1988bq,Beenakker:1990ma}.  Numerically, this
modification of $\Delta c_1$ turned out to be negligible.
(ii) In a publication by two of us \cite{Kramer:2003cw}, subtraction
terms for the non-Abelian part of the process $\gamma g \to Q\Qbar g$
have been derived by comparing the zero-mass limit of Ref.\ 
\cite{Merebashvili:2000ya} with the massless theory of Ref.\ 
\cite{Gordon:1994wu}, which do not meet the expectations of mass
factorization in Sec.\ \ref{sec:convolution}.  The subtraction terms
derived this way correctly describe the transition between the two
theories. Obviously, if one of the theories uses conventions differing
from the conventional $\msbar$ scheme, the results will not agree with
the subtraction terms derived via mass factorization.  Whether this is
indeed the reason for the discrepancy, can be clarified only with the
help of a new full calculation.  It is noteworthy that also for the
channel $\gamma q \to Q\Qbar q$ a non-vanishing result for the
coefficient $\Delta c_{11}$ (see Eq.\ (78) in Ref.\ 
\cite{Kramer:2003cw}) was found, which would have been zero employing
the methods in Sec.\ \ref{sec:convolution}.  In this case, the
difference could be traced back to an error in Ref.\ 
\cite{Gordon:1994wu}.

 
\section*{Acknowledgment} 

We are grateful to I.\ Bojak for providing his {\tt FORTRAN} code for
the hadroproduction of heavy flavors at NLO, which made this work
possible. We would like to thank S.\ Kretzer and F.\ Olness for useful
discussions.  I.\ S.\ is grateful to the theory group at Fermilab for
the kind hospitality extended to him last summer. During this visit, the
present paper was initiated.  The work of I.\ S.\ was supported by DESY.
This work was supported in part by the Bundesministerium f\"ur Bildung
und Forschung through Grant No.\ 05 HT4GUA/4.


\appendix

\section{Cross sections for $2 \to 2$ subprocesses}
\label{app:subprocesses}

In this appendix, we list the cross sections for all
one-particle-inclusive subprocesses, $a+b\to c+X$, needed to compute the
subtraction terms in Sec.\ \ref{sec:subtraction}.  For brevity, part $X$
of the final state is not written explicitly in the following.  We begin
with subprocesses occurring in the channel $\ggQ$, needed in Sec.\ 
\ref{sec:cutgg}.


\subsection{Subprocesses in $\ggQ$}
\label{app:cutgg}

\begin{eqnarray}
\frac{\der \sigmahat^{(0)}}{\der \vhat}(g g \to Q) &=&
\alpha_s^2 \pi \frac{1}{(N^2-1)} \frac{1}{\shat} 
\left[C_F - N \vhat (1-\vhat)\right] \tau(\shat,\vhat) \, ,
\\
\nonumber\\
\frac{\der \sigmahat^{(0)}}{\der \vhat}(g g \to g) &=&
\alpha_s^2 \pi \frac{4 N^2}{N^2-1} \frac{1}{\shat} 
\frac{(1-x)^3}{x^2}\, ,
\\
\nonumber\\
\frac{\der \sigmahat^{(0)}}{\der \vhat}(g Q \to Q) &=&
\alpha_s^2 \pi \frac{1}{N^2-1} \frac{1}{\shat} \frac{1 + (1-\vhat)^2}{\vhat}
2 C_F 
[C_F \vhat^2 + N (1-\vhat)] \frac{1}{\vhat (1-\vhat)} \, ,
\\
\nonumber\\
\frac{\der \sigmahat^{(0)}}{\der \vhat}(Q g \to Q) &=&
\left.
\frac{\der \sigmahat^{(0)}}{\der \vhat}(g Q\to Q)
\right|_{\vhat \leftrightarrow 1-\vhat} 
\nonumber\\
&=&
\alpha_s^2 \pi \frac{1}{N^2-1} \frac{1}{\shat} \frac{1 + \vhat^2}{1-\vhat}
2 C_F 
\left[C_F (1-\vhat)^2 + N \vhat\right] \frac{1}{\vhat (1-\vhat)} \, ,
\end{eqnarray}
where
\begin{eqnarray}
\tau(\shat,\vhat) & = &
\frac{\vhat}{1-\vhat} + \frac{1-\vhat}{\vhat}
+ \frac{4 m^2}{\shat \vhat (1-\vhat)} 
\left(1-\frac{m^2}{\shat \vhat (1-\vhat)}\right) \, ,
\\[1.5ex]
x & = & \vhat (1-\vhat) \, . 
\end{eqnarray}


\subsection{Subprocesses in $\qqbQ$}
\label{app:cutqq}

\begin{eqnarray}
\frac{\der \sigmahat^{(0)}}{\der \vhat}(q \qbar \to Q) &=&
\alpha_s^2 \pi \frac{C_F}{N} \frac{1}{\shat} 
\left[ 
(1-\vhat)^2 + \vhat^2 + \frac{2 m^2}{\shat}
\right] \, ,
\\
\nonumber\\
\frac{\der \sigmahat^{(0)}}{\der \vhat}(q \qbar \to g) &=&
\left.\frac{\der \sigmahat^{(0)}}{\der \vhat}(g g \to Q)\right|_{m \to 0}
\nonumber\\
&=&
\alpha_s^2 \pi \frac{1}{N^2-1} \frac{1}{\shat} 
\left[C_F - N \vhat (1-\vhat)\right] 
\left(\frac{\vhat}{1-\vhat} + \frac{1-\vhat}{\vhat}\right) \, .
\end{eqnarray}


\subsection{Subprocesses in $\gqQ$ and $\gqbQ$}
\label{app:cutgq}

\begin{eqnarray}
\frac{\der \sigmahat^{(0)}}{\der \vhat}(q g \to g) &=&
\alpha_s^2 \pi \frac{1}{2 N^2} \frac{1}{\shat} (2- 2 \vhat + \vhat^2)
\left[(N^2-1) \vhat^2 + 2 N^2 (1-\vhat)\right] 
\frac{1}{\vhat^2 (1-\vhat)} \, ,
\\
\nonumber\\
\frac{\der \sigmahat^{(0)}}{\der \vhat}(gq \to g) &=&
\left.
\frac{\der \sigmahat^{(0)}}{\der \vhat}(q g \to g)
\right|_{\vhat \leftrightarrow 1-\vhat}
\nonumber\\
&=&
\alpha_s^2 \pi \frac{1}{2 N^2} \frac{1}{\shat} (1+ \vhat^2)
\left[(N^2-1) \vhat^2 +2 \vhat + (N^2-1)\right] 
\frac{1}{\vhat (1-\vhat)^2} \, ,
\\
\nonumber\\
\frac{\der \sigmahat^{(0)}}{\der \vhat}(Q q  \to Q) &=&
\alpha_s^2 \pi \frac{C_F}{N} \frac{1}{\shat} 
\frac{1 + \vhat^2}{(1-\vhat)^2} \, ,
\\
\nonumber\\
\frac{\der \sigmahat^{(0)}}{\der \vhat}(Q \qbar \to Q) &=&
\frac{\der \sigmahat^{(0)}}{\der \vhat}(Q q \to Q) \, .
\end{eqnarray}


\section{Feynman diagrams}
\label{app:feynman}

In this appendix we list the bremsstrahlung Feynman diagrams
contributing at NLO to the process $p + \pbar \to H + X$ ($H$ denotes a
heavy meson, $D$, $D^\star$, $B\ldots$) in the FFNS.  They are the basis
to generate the cut diagrams Figs.\ \ref{fig:cutgg1}--\ref{fig:cutgq2}
as described in Sec.\ \ref{sec:subatnlo}.  We show separately the subset
of Feynman diagrams for $gg \to Q \Qbar g$ which, after replacing one of
the incoming gluons by a photon, contribute also to heavy-quark
photoproduction, $\gamma + p \to H + X$.

\begin{figure}[htb]
\begin{center}
\setlength{\unitlength}{1pt}
\begin{picture}(342,185)
\put(0,130){
  \begin{picture}(102,63)
    \SetWidth{0.7}
    \Photon(0,56)(51,56){3}{4}
    \Gluon(0,7)(51,7){4}{5}
    \Gluon(70,56)(102,31.5){-3}{4}
    \Vertex(51,56){2.0}
    \Vertex(51,7){2.0}
    \Vertex(70,56){2.0}
    \SetWidth{1.8}
    \ArrowLine(102,7)(51,7)
    \ArrowLine(51,7)(51,56)
    \Line(51,56)(76.5,56)
    \ArrowLine(76.5,56)(102,56)
  \end{picture}
}
\Text(55,110)[a]{\bf (a)}
\put(120,130){
  \begin{picture}(102,63)
    \SetWidth{0.7}
    \Photon(0,56)(51,56){3}{4}
    \Gluon(0,7)(51,7){4}{5}
    \Gluon(51,31.5)(102,31.5){3}{6}
    \Vertex(51,56){2.0}
    \Vertex(51,7){2.0}
    \Vertex(51,31.5){2.0}
    \SetWidth{1.8}
    \ArrowLine(102,7)(51,7)
    \Line(51,7)(51,31.5)
    \ArrowLine(51,31.5)(51,56)
    \ArrowLine(51,56)(102,56)
  \end{picture}
}
\Text(175,110)[b]{\bf (b)}
\put(240,130){
  \begin{picture}(102,63)
    \SetWidth{0.7}
    \Photon(0,56)(51,56){3}{4}
    \Gluon(0,7)(51,7){4}{5}
    \Gluon(70,7)(102,31.5){3}{4}
    \Vertex(51,56){2.0}
    \Vertex(51,7){2.0}
    \Vertex(70,7){2.0}
    \SetWidth{1.8}
    \ArrowLine(102,7)(76.5,7)
    \Line(76.5,7)(51,7)
    \ArrowLine(51,7)(51,56)
    \ArrowLine(51,56)(102,56)
  \end{picture}
}
\Text(295,110)[b]{\bf (c)}
\put(0,20){
  \begin{picture}(102,63)
    \SetWidth{0.7}
    \Photon(0,56)(51,56){3}{4}
    \Gluon(0,7)(51,7){3}{7}
    \Gluon(51,7)(102,7){3}{7}
    \Line(51,7)(51,12)
    \Gluon(51,12)(51,31.5){3}{3}
    \Vertex(51,56){2.0}
    \Vertex(51,7){2.0}
    \Vertex(51,31.5){2.0}
    \SetWidth{1.8}
    \ArrowLine(102,31.5)(51,31.5)
    \ArrowLine(51,31.5)(51,56)
    \ArrowLine(51,56)(102,56)
  \end{picture}
}
\Text(55,0)[b]{\bf (d)}
\end{picture}
\end{center}
\caption{Feynman diagrams for the NLO gluon bremsstrahlung process
  $\gamgQ$. Reversing the heavy-quark lines yields the remaining graphs.
  Diagrams obtained from the ones shown here by replacing the photon
  with a gluon contribute to $\ggQ$.  }
\label{fig:gamg}
\end{figure}
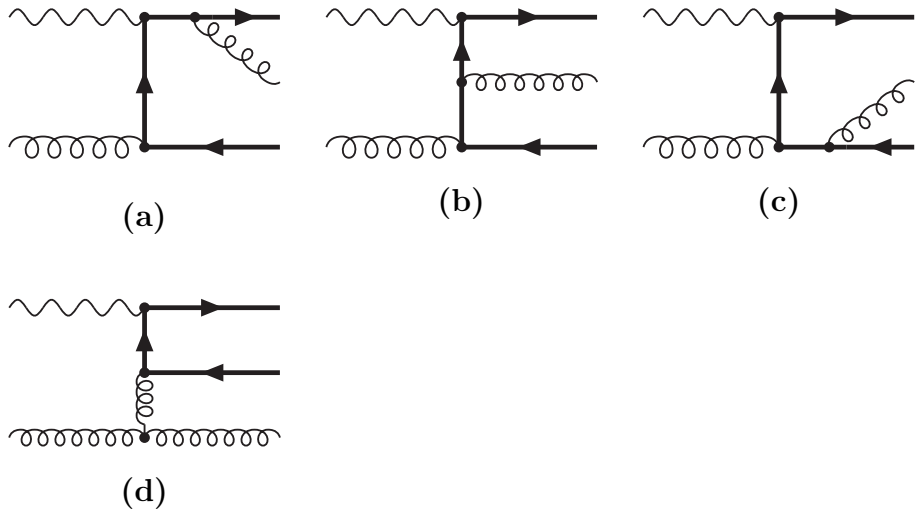

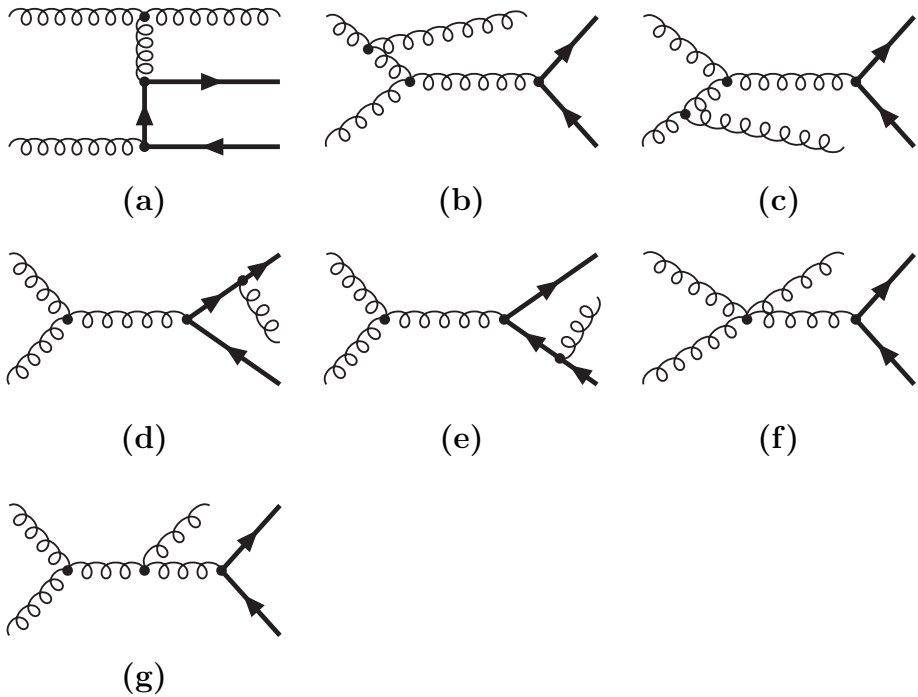
\begin{figure}[ht!]
\begin{center}
\setlength{\unitlength}{1pt}
\begin{picture}(342,270)
\put(0,190){
  \begin{picture}(102,63)
    \SetWidth{0.7}
    \Gluon(0,7)(51,7){3}{7}
    \Gluon(0,56)(51,56){3}{7}
    \Gluon(51,56)(102,56){3}{7}
   \Gluon(51,31.5)(51,56){3}{4}
    \Vertex(51,56){2.0}
    \Vertex(51,31.5){2.0}
    \Vertex(51,7){2.0}
    \SetWidth{1.8}
    \ArrowLine(51,31.5)(102,31.5)
    \ArrowLine(51,7)(51,31.5)
    \ArrowLine(102,7)(51,7)
  \end{picture}
}
\Text(55,170)[b]{\bf (a)}
\put(120,190){
  \begin{picture}(102,63)
    \SetWidth{0.7}
    \Gluon(0,56)(15.75,43.75){3}{2}
    \Gluon(15.75,43.75)(31.5,31.5){3}{2}
    \Gluon(0,7)(31.5,31.5){3}{4}
    \Gluon(31.5,31.5)(80,31.5){3}{6}
    \Gluon(15.75,43.75)(75.5,56){3}{7}
    \Vertex(15.75,43.75){2.0}
    \Vertex(31.5,31.5){2.0}
    \Vertex(80,31.5){2.0}
    \SetWidth{1.8}
    \ArrowLine(102,7)(80,31.5)
    \ArrowLine(80,31.5)(102,56)
  \end{picture}
}
\Text(175,170)[b]{\bf (b)}
\put(240,190){
  \begin{picture}(102,63)
    \SetWidth{0.7}
    \Gluon(0,56)(31.5,31.5){3}{4}
    \Gluon(15.75,19.25)(31.5,31.5){3}{2}
    \Gluon(0,7)(15.75,19.25){3}{2}
    \Gluon(31.5,31.5)(80,31.5){3}{6}
    \Gluon(75.5,7)(15.75,19.25){3}{7}
    \Vertex(15.75,19.25){2.0}
    \Vertex(31.5,31.5){2.0}
    \Vertex(80,31.5){2.0}
    \SetWidth{1.8}
    \ArrowLine(102,7)(80,31.5)
    \ArrowLine(80,31.5)(102,56)
  \end{picture}
}
\Text(295,170)[b]{\bf (c)}
\put(0,100){
  \begin{picture}(102,63)
    \SetWidth{0.7}
    \Gluon(0,56)(22,31.5){3}{4}
    \Gluon(0,7)(22,31.5){3}{4}
    \Gluon(22,31.5)(67,31.5){3}{5}
    \Gluon(102,23)(88,46.2){3}{3}
    \Vertex(22,31.5){2.0}
    \Vertex(67,31.5){2.0}
    \Vertex(88,46.2){2.0}
    \SetWidth{1.8}
    \ArrowLine(102,7)(67,31.5)
    \ArrowLine(88,46.2)(102,56)
    \ArrowLine(67,31.5)(88,46.2)
  \end{picture}
}
\Text(55,80)[b]{\bf (d)}
\put(120,100){
  \begin{picture}(102,63)
    \SetWidth{0.7}
    \Gluon(0,56)(22,31.5){3}{4}
    \Gluon(0,7)(22,31.5){3}{4}
    \Gluon(22,31.5)(67,31.5){3}{5}
    \Gluon(102,40)(88,16.8){3}{3}
    \Vertex(88,16.8){2.0}
    \Vertex(22,31.5){2.0}
    \Vertex(67,31.5){2.0}
    \SetWidth{1.8}
    \ArrowLine(102,7)(88,16.8)
    \ArrowLine(88,16.8)(67,31.5)
    \ArrowLine(67,31.5)(102,56)
  \end{picture}
}
\Text(175,80)[b]{\bf (e)}
\put(240,100){
  \begin{picture}(102,63)
    \SetWidth{0.7}
    \Gluon(0,56)(39,31.5){3}{5}
    \Gluon(0,7)(39,31.5){3}{5}
    \Gluon(39,31.5)(80,31.5){3}{4}
    \Gluon(39,31.5)(75.5,56){3}{4}
    \Vertex(39,31.5){2.0}
    \Vertex(80,31.5){2.0}
    \SetWidth{1.8}
    \ArrowLine(102,7)(80,31.5)
    \ArrowLine(80,31.5)(102,56)
  \end{picture}
}
\Text(295,80)[b]{\bf (f)}
\put(0,5){
  \begin{picture}(102,63)
    \SetWidth{0.7}
    \Gluon(0,56)(22,31.5){3}{4}
    \Gluon(0,7)(22,31.5){3}{4}
    \Gluon(22,31.5)(51,31.5){3}{3}
    \Gluon(51,31.5)(80,31.5){3}{3}
    \Line(51,31.5)(51,33.5)
    \Gluon(51,33.5)(75.5,56){3}{3}
    \Vertex(22,31.5){2.0}
    \Vertex(51,31.5){2.0}
    \Vertex(80,31.5){2.0}
    \SetWidth{1.8}
    \ArrowLine(102,7)(80,31.5)
    \ArrowLine(80,31.5)(102,56)
  \end{picture}
}
\Text(55,-10)[b]{\bf (g)}
\end{picture}
\end{center}
\caption{Additional Feynman diagrams for the NLO gluon bremsstrahlung
  process $\ggQ$.  Replacing the photons by gluons in Fig.\ 
  \protect\ref{fig:gamg} and reversing the heavy-quark lines of part (a)
  yields the remaining graphs.}
\label{fig:gg}
\end{figure}

\begin{figure}[th!]
\begin{center}
\setlength{\unitlength}{1pt}
\begin{picture}(342,160)
\put(0,100){
  \begin{picture}(102,63)
    \SetWidth{0.7}
    \ArrowLine(0,56)(15.75,43.75)
    \ArrowLine(15.75,43.75)(31.5,31.5)
    \ArrowLine(31.5,31.5)(0,7)
    \Gluon(31.5,31.5)(80,31.5){3}{6}
    \Gluon(15.75,43.75)(75.5,56){3}{7}
    \Vertex(15.75,43.75){2.0}
    \Vertex(31.5,31.5){2.0}
    \Vertex(80,31.5){2.0}
    \SetWidth{1.8}
    \ArrowLine(102,7)(80,31.5)
    \ArrowLine(80,31.5)(102,56)
  \end{picture}
}
\Text(55,80)[a]{\bf (a)}
\put(120,100){
  \begin{picture}(102,63)
    \SetWidth{0.7}
    \ArrowLine(0,56)(31.5,31.5)
    \ArrowLine(31.5,31.5)(15.75,19.25)
    \ArrowLine(15.75,19.25)(0,7)
    \Gluon(31.5,31.5)(80,31.5){3}{6}
    \Gluon(75.5,7)(15.75,19.25){3}{7}
    \Vertex(15.75,19.25){2.0}
    \Vertex(31.5,31.5){2.0}
    \Vertex(80,31.5){2.0}
    \SetWidth{1.8}
    \ArrowLine(102,7)(80,31.5)
    \ArrowLine(80,31.5)(102,56)
  \end{picture}
}
\Text(175,80)[b]{\bf (b)}
\put(240,100){
  \begin{picture}(102,63)
    \SetWidth{0.7}
    \ArrowLine(0,56)(22,31.5)
    \ArrowLine(22,31.5)(0,7)
    \Gluon(22,31.5)(67,31.5){3}{5}
    \Gluon(102,23)(88,46.2){3}{3}
    \Vertex(22,31.5){2.0}
    \Vertex(67,31.5){2.0}
    \Vertex(88,46.2){2.0}
    \SetWidth{1.8}
    \ArrowLine(102,7)(67,31.5)
    \ArrowLine(88,46.2)(102,56)
    \ArrowLine(67,31.5)(88,46.2)
  \end{picture}
}
\Text(295,80)[b]{\bf (c)}
\put(57,10){
  \begin{picture}(102,63)
    \SetWidth{0.7}
    \ArrowLine(0,56)(22,31.5)
    \ArrowLine(22,31.5)(0,7)
    \Gluon(22,31.5)(67,31.5){3}{5}
    \Gluon(102,40)(88,16.8){3}{3}
    \Vertex(22,31.5){2.0}
    \Vertex(67,31.5){2.0}
    \Vertex(88,16.8){2.0}
    \SetWidth{1.8}
    \ArrowLine(102,7)(88,16.8)
    \ArrowLine(88,16.8)(67,31.5)
    \ArrowLine(67,31.5)(102,56)
  \end{picture}
}
\Text(108,-10)[b]{\bf (d)}
\put(177,10){
  \begin{picture}(102,63)
    \SetWidth{0.7}
    \ArrowLine(0,56)(22,31.5)
    \ArrowLine(22,31.5)(0,7)
    \Gluon(22,31.5)(51,31.5){3}{3}
    \Gluon(51,31.5)(80,31.5){3}{3}
    \Line(51,31.5)(51,33.5)
    \Gluon(51,33.5)(75.5,56){3}{3}
    \Vertex(22,31.5){2.0}
    \Vertex(51,31.5){2.0}
    \Vertex(80,31.5){2.0}
    \SetWidth{1.8}
    \ArrowLine(102,7)(80,31.5)
    \ArrowLine(80,31.5)(102,56)
  \end{picture}
}
\Text(228,-10)[b]{\bf (e)}
\end{picture}
\end{center}
\caption{Feynman diagrams for the NLO gluon bremsstrahlung process
  $\qqbQ$.} 
\end{figure}
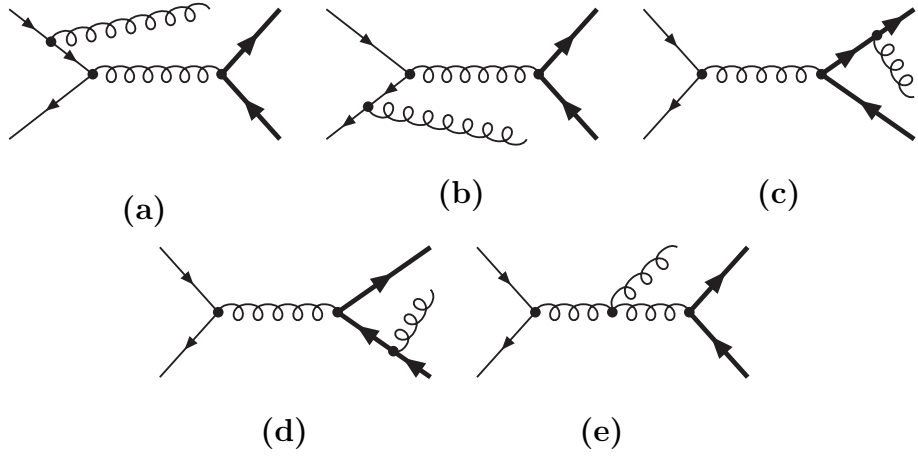
\begin{figure}[ht!]
\begin{center}
\setlength{\unitlength}{1pt}
\begin{picture}(342,180)
\put(51,100){
  \begin{picture}(102,63)
    \SetWidth{0.7}
    \Photon(0,56)(51,56){3}{4}
    \ArrowLine(0,7)(51,7)
    \ArrowLine(51,7)(102,7)
    \Gluon(51,7)(51,31.5){3}{3}
    \Vertex(51,56){2.0}
    \Vertex(51,7){2.0}
    \Vertex(51,31.5){2.0}
    \SetWidth{1.8}
    \ArrowLine(102,31.5)(51,31.5)
    \ArrowLine(51,31.5)(51,56)
    \ArrowLine(51,56)(102,56)
  \end{picture}
}
\Text(102,80)[b]{\bf (a)}
\put(189,100){
  \begin{picture}(102,63)
    \SetWidth{0.7}
    \Photon(0,56)(51,56){3}{4}
    \ArrowLine(0,7)(51,7)
    \ArrowLine(51,7)(102,7)
    \Gluon(51,7)(51,31.5){3}{3}
    \Vertex(51,56){2.0}
    \Vertex(51,7){2.0}
    \Vertex(51,31.5){2.0}
    \SetWidth{1.8}
    \ArrowLine(51,31.5)(102,31.5)
    \ArrowLine(51,56)(51,31.5)
    \ArrowLine(102,56)(51,56)
  \end{picture}
}
\Text(240,80)[b]{\bf (b)}
\put(51,10){
  \begin{picture}(102,63)
    \SetWidth{0.7}
    \Photon(0,56)(32,20){3}{4}
    \Gluon(70,20)(80,43.75){3}{3}
    \ArrowLine(0,7)(32,20)
    \ArrowLine(32,20)(70,20)
    \ArrowLine(70,20)(102,7)
    \Vertex(32,20){2.0}
    \Vertex(70,20){2.0}
    \Vertex(80,43.75){2.0}
    \SetWidth{1.8}
    \ArrowLine(102,31.5)(80,43.75)
    \ArrowLine(80,43.75)(102,56)
  \end{picture}
}
\Text(102,-10)[b]{\bf (c)}
\put(189,10){
  \begin{picture}(102,63)
    \SetWidth{0.7}
    \Photon(0,56)(44.7,33){3}{4}
    \Photon(54.4,28)(70,20){-3}{1.3}
    \Gluon(32,20)(80,43.75){3}{5}
    \ArrowLine(0,7)(32,20)
    \ArrowLine(32,20)(70,20)
    \ArrowLine(70,20)(102,7)
    \Vertex(32,20){2.0}
    \Vertex(70,20){2.0}
    \Vertex(80,43.75){2.0}
    \SetWidth{1.8}
    \ArrowLine(102,31.5)(80,43.75)
    \ArrowLine(80,43.75)(102,56)
  \end{picture}
}
\Text(240,-10)[b]{\bf (d)}
\end{picture}
\end{center}
\caption{Feynman diagrams for the  NLO light-quark-initiated subprocess
  $\gamqQ$: ``Bethe-Heitler'' graphs (a) and (b), ``Compton'' graphs (c)
  and (d). Diagrams obtained from the ones shown here by replacing the
  photon with a gluon contribute to $\gqQ$.}
\label{fig:gamq}
\end{figure}
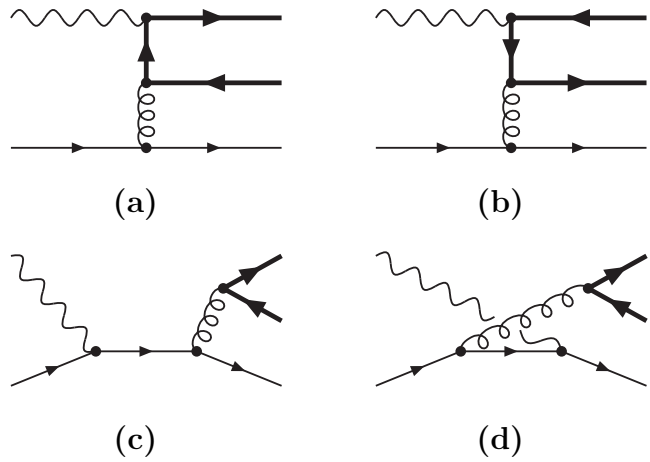
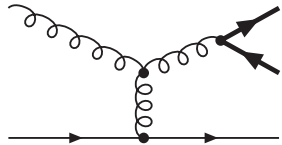
\begin{figure}[htb!]
\begin{center}
\setlength{\unitlength}{1pt}
\begin{picture}(102,49)
\put(0,-10){
\begin{picture}(102,55)
  \SetWidth{0.7}
  \Gluon(0,56)(51,31.5){3}{5}
  \ArrowLine(0,7)(51,7)
  \ArrowLine(51,7)(102,7)
  \Gluon(51,7)(51,31.5){3}{3}
  \Gluon(51,31.5)(80,43.75){3}{3}
  \Vertex(51,7){2.0}
  \Vertex(51,31.5){2.0}
  \Vertex(80,43.75){2.0}
  \SetWidth{1.8}
  \ArrowLine(102,31.5)(80,43.75)
  \ArrowLine(80,43.75)(102,56)
\end{picture}
}
\end{picture}
\end{center}
\caption{Additional Feynman diagram for the NLO light-quark-initiated
  subprocess $\gqQ$. Replacing the photons by gluons in Fig.\
  \protect\ref{fig:gamq} yields the remaining graphs.} 
\label{fig:gq}
\end{figure}


\end{document}